\documentclass[aps,prl,twocolumn,nofootinbib,twocolumn,superscriptaddress]{revtex4-1}

\usepackage{graphicx}
\usepackage{dcolumn}
\usepackage{bm}
\usepackage{ textcomp }
\usepackage{color}
\usepackage{amsmath}
\usepackage{amssymb}
\usepackage{xcolor}
\usepackage{hyperref}
\usepackage{easyReview}
\usepackage{mathdots}
\usepackage{float}
\usepackage{lineno}
\usepackage{array}
\usepackage{xcolor,colortbl}
\usepackage[normalem]{ulem}
\definecolor{LightBlue}{rgb}{0.8,0.8,0.8}

\usepackage{amsmath} 

\newcommand{\eph}{\text{\it e}-ph}
\allowdisplaybreaks

\begin{document}
\title{Topological Superconductivity From Forward Phonon Scatterings}

\author{Shaozhi Li}
\email[]{lishaozhiphys@gmail.com}
\affiliation{Materials Science and Technology Division, Oak Ridge National Laboratory, Oak Ridge, Tennessee 37831, USA}

\author{Lun-Hui Hu}
\affiliation{Department of Physics and Astronomy, The University of Tennessee, Knoxville, Tennessee 37966, USA}
\affiliation{Institute for Advanced Materials and Manufacturing, University of Tennessee, Knoxville, Tennessee 37920, USA}

\author{Rui-Xing Zhang}
\affiliation{Department of Physics and Astronomy, The University of Tennessee, Knoxville, Tennessee 37966, USA}
\affiliation{Institute for Advanced Materials and Manufacturing, University of Tennessee, Knoxville, Tennessee 37920, USA}
\affiliation{Department of Materials Science and Engineering, The University of Tennessee, Knoxville, Tennessee 37966, USA}

\author{Satoshi Okamoto}
\affiliation{Materials Science and Technology Division, Oak Ridge National Laboratory, Oak Ridge, Tennessee 37831, USA}

\date{\today}

\begin{abstract}
We propose a new Rashba-free mechanism to realize topological superconductivity with electron-phonon interaction. In the presence of a magnetic field, electron-phonon interaction with small momentum transfer is found to favor spin-triplet Cooper pairing. This process facilitates the formation of chiral topological superconductivity even when Rashba spin-orbital coupling is absent. As a proof of concept, we propose an experimentally feasible heterostructure to systematically study the entangled relationship among forward-phonon scatterings, Rashba spin-orbital couplings, pairing symmetries, and superconducting topology. Our theory sheds light on the important role of electron-phonon coupled materials in the pursuit of non-Abelian Majorana quasiparticles. 
\end{abstract}

\maketitle
{\it Introduction - } 
Topological superconductors have been attracting broad interests due to the great potential application of the non-Abelian states in fault-tolerant quantum computing~\cite{FreedmanPRB2006,Nayak2008,Field_2018,BiaoPNAS2018,Lutchyn}.
Topological physics in superconductors can arise in an intrinsic or extrinsic manner~\cite{McM1968,KanePRL2008,QiPRL2009,Alicea_2012,Beenakker2013,Jiang_2013,XuPRL2014,Sajadi2018,ZhangScience2018,WangScience2018,Lee2019PRB,Machida2019,Zhuscience2019,RameauPRB2019,Trang2020,Wangscience2020,ChengNP2020,Mascot2021,ZhangPRL2021}. For example, unconventional superconductors with chiral $p$-wave Cooper pairing, are a promising platform to achieve topological superconductivity intrinsically in a bulk system~\cite{QiRevModPhys,Jirscience2021,ChouPRL2021}. Because unconventional superconductors are scarce in nature, recent research attentions have been focusing on extrinsic approaches to achieve ``artificial" topological superconductor via hetero-engineering~\cite{McM1968,KanePRL2008,QiPRL2009,Alicea_2012,Beenakker2013,Jiang_2013,XuPRL2014,Trang2020}. These artificial topological superconductors usually consist of a conventionally superconducting substrate as a Cooper pair source and a functional add-on layer with a strong spin-orbital coupling. Choice of the add-on layer could be a Rashba system or a topological insulator surface, and it needs to borrow Cooper pairs from the substrate, as well as an applied magnetic field, to make topological superconductor. Thanks to the state-of-the-art epitaxial growth technique, recent years have witnessed significant progress along the direction of extrinsic topological superconductors~\cite{HorPRL2010,LiangPRL2010,Mourik2012,Kallin_2012,Das2012,GangPRL2016,Liunpj2017,LiNC2021}, including Rashba nanowires~\cite{Mourik2012}, ferromagnetic semiconductors~\cite{KezilebiekeNature}, and high-mobility two-dimensional (2D) electron gas~\cite{Mortezaarxiv}. However, superconductivity from the proximity effect is usually weaker than an intrinsic one, significantly limiting the working temperature of artificial topological superconductor setups. Furthermore, a small superconducting gap of topological superconductors will necessarily increase the vulnerability of its topological boundary modes (i.e., Majorana modes) regarding disordering effects~\cite{HuiPRB2015,YaoPRB2020}. Therefore, an interesting open question in this field is how to break this temperature barrier of artificial topological superconductor in heterostructures.

\begin{figure}[t]
\center\includegraphics[width=\columnwidth]{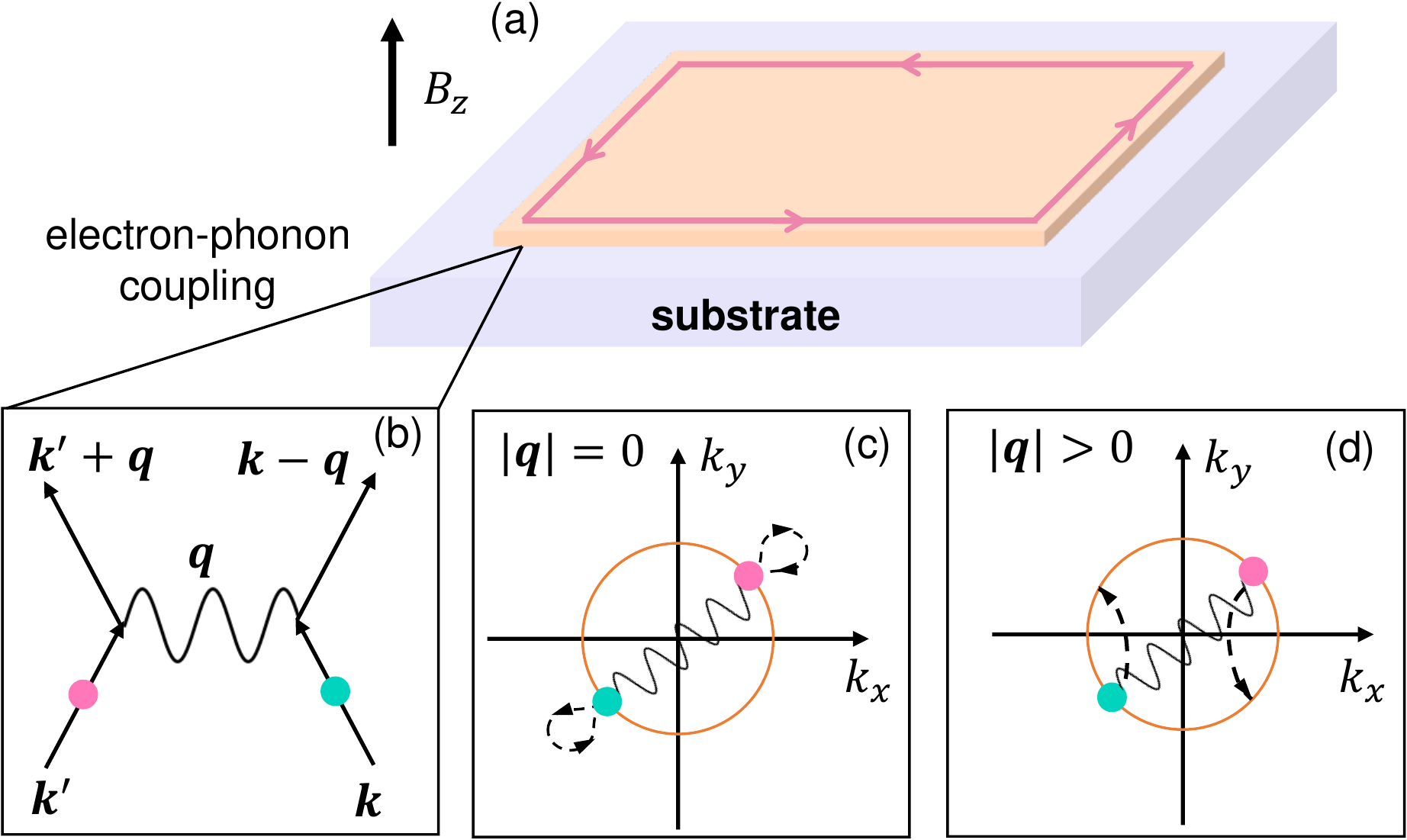}
\caption{\label{Fig:fig0} Schematic view of the 2D topological superconductor induced by the interfacial electron-phonon interaction. Panel~(b) shows two-electron scattering by a phonon with the momentum transfer $\bf q$. 
Panel~(c) shows electron scattering near the Fermi level, relevant to the paring, with zero momentum transfer $\bf q=0$. Panel~(d) is same as panel~(c) but with a nonzero momentum transfer $\bf q$.}
\end{figure}

Our main finding in this work is that {\it forward electron-phonon scattering tends to generate spin-triplet Cooper pairs in the presence of a magnetic field}, which motivates us to propose a new experimentally feasible paradigm to achieve high-temperature artificial chiral topological superconductor, without using the superconducting proximity effect~\cite{Tudor2022}. As shown in Fig.~\ref{Fig:fig0}, the heterostructure we considered involves a nonsuperconducting substrate that generates phonons, instead of Cooper pairs, to a 2D add-on layer, which could have Rashba spin-orbital coupling (RSOC). The interaction between electrons and substrate phonons leads to Bardeen-Cooper-Schrieffer (BCS) instability, further generating {\it intrinsic} superconductivity in the add-on layer. 
Previous studies have shown that this interfacial interaction driven superconductivity can exist at high temperatures~\cite{LeeNature2014,Lee_2015,Rademaker_2016,ZhangPRB2016,RebecPRL2017,SongNC2019,RademakerPRB2021}, implying a pathway to avoid the temperature limitation of the superconducting proximity effect.
Remarkably, we find that short-ranged forward phonon scatterings are crucial to induce triplet pairing, and further significantly stabilizes superconductivity against an applied magnetic field. This effect allows our setup to go beyond the Pauli limit~\cite{Chandrasekhar1962,ClogstonPRL1962,XiePRL2020} and further fulfill the stringent topological superconducting condition for RSOC systems~\cite{sauPRL2010}.

Unlike most proposals for artificial topological superconductors in the literature, RSOC is {\it not} a necessary ingredient for our phonon-mediated topological superconductor recipe. In the absence of RSOC, we find that the magnetic field drives a phase transition from a spin-singlet pairing phase to a triplet one, allowing the Zeeman-induced spin-polarized Fermi surface to be Cooper paired. This phenomenon is crucial for enabling a zero-RSOC topological superconductor. 
When RSOC is turned on, the breaking of inversion symmetry leads to a singlet-triplet mixed state, while the external magnetic field driven singlet-triplet transition now becomes a crossover (e.g.,~singlet-dominant to triplet-dominant pairings).



{\it Model Hamiltonian} -  
We first set up a platform for a two-dimensional single-band electron gas with {\eph} coupling as shown in Fig.~\ref{Fig:fig0}. The entire Hamiltonian of this system is given by $H=H_0+H_\text{ph}+H_\text{e-ph}$, where
\begin{widetext}
 \begin{eqnarray}
H_0=-2t\sum_{{\bf k},\sigma} \left[\text{cos}(k_x a)+\text{cos}(k_y a) \right] c_{{\bf k},\sigma}^{\dagger}c_{{\bf k},\sigma}^{\phantom\dagger}
+ 2\lambda_\text{SOC} \sum_{{\bf k}}\left[ d({\bf k})c_{{\bf k},\uparrow}^{\dagger}c_{{\bf k},\downarrow}^{\phantom\dagger} + h.c.\right]  + \sum_{{\bf k},\sigma}(\delta_{\sigma} B_z-\mu)c_{{\bf k},\sigma}^{\dagger}c_{{\bf k},\sigma}^{\phantom\dagger},
\end{eqnarray}
\end{widetext}
\begin{eqnarray}
H_\text{ph}&=&\Omega\sum_{{\bf q}}b_{\bf q}^{\dagger}b_{\bf q}^{\phantom\dagger},\\ 
H_\text{e-ph}&=&\frac{1}{\sqrt{N}}\sum_{{\bf k},{\bf q},\sigma}g({\bf q})c_{{\bf k}+{\bf q},\sigma}^{\dagger}c_{{\bf k},\sigma}^{\phantom\dagger}(b_{-{\bf q}}^{\dagger}+b_{{\bf q}}^{\phantom\dagger}).
\end{eqnarray}
Here, $c_{{\bf k},\sigma}^{\dagger}$ ($c_{{\bf k},\sigma}^{\phantom\dagger}$) creates (annihilates) a spin $\sigma$ ($=\uparrow,\downarrow$) electron with momentum ${\bf k}$. $b_{{\bf q}}^{\dagger}$ ($b_{\bf q}^{\phantom\dagger}$) creates (annihilates) a phonon mode with momentum ${\bf q}$. $d({\bf k})=\sin (k_ya)+i \sin (k_xa)$, where $a$ is the lattice constant. $t$ is the nearest neighbor hopping integral, $\lambda_\text{SOC}$ is the Rashba spin-orbit coupling strength, $\Omega$ is the optical phonon frequency, $B_z$ is the magnetic field, $\mu$ is the chemical potential, and $N$ is the lattice size. $\delta_{\sigma}=-1$ (1) for spin-up (spin-down) electrons. $g({\bf q})$ is the matrix element of the {\eph} interaction, and we adopt $g^2({\bf q})=g^2_0f({\bf q})$. Motivated by the {\eph} coupling physics at the interface of FeSe/STO~\cite{Rademaker_2016,RademakerPRB2021}, both uniform and forward {\eph} couplings are studied in this work. While $f({\bf q})=1$ denotes a uniform {\eph} interaction, the forward {\eph} interaction is captured by a Gaussian-like form factor $f({\bf q})=4\pi^2\frac{1}{\sqrt{2\pi}q_0} e^{-|{\bf q}|/q_0}$, where $q_0$ is the scattering range.

We analyze this Hamiltonian using the Migdal-Eliashberg (ME) theory in the Nambu space~\cite{MARSIGLIO2020168102,supple}. Throughout this work, we set the lattice size $N=40\times 40$, the temperature $T=0.01t$, the dimensionless {\eph} coupling strength $\lambda_{ph}=\frac{2g_0^2}{W\Omega}\langle f({\bf q})\rangle =0.8$, and the electron density $n=0.2$. A weak spin-orbit coupling has a tiny effect on the bandwidth $W$, and we set $W=8t$ here. We follow Ref.~\cite{Rademaker_2016} and assume that the relevant phonon branch for the forward scattering is an optical phonon with $\Omega=t$.
Stability of our conclusion against a scaling of $\lambda_{ph}$ has also been tested and confirmed in the supplemental material~\cite{supple}. To further extract the topological information, an effective single-particle Hamiltonian is constructed to fit the ME theory, with $H^{\prime}=H_0+\sum_{{\bf k}}\Sigma({\bf k})$. In the Nambu space, the self-energy at zero frequency $\Sigma({\bf k})$ is a $4\times 4$ matrix, obtained by extrapolating the self-energy $\Sigma({\bf k},i\omega_n)$ in the Matsubara frequency $\omega_n$ space as $\omega_n \rightarrow 0$. Topological invariant such as the first Chern number of $H^\prime$ can thus be calculated in an efficient way~\cite{FukuiJPSJ2005}.

\begin{figure}[t]
\center\includegraphics[width=\columnwidth]{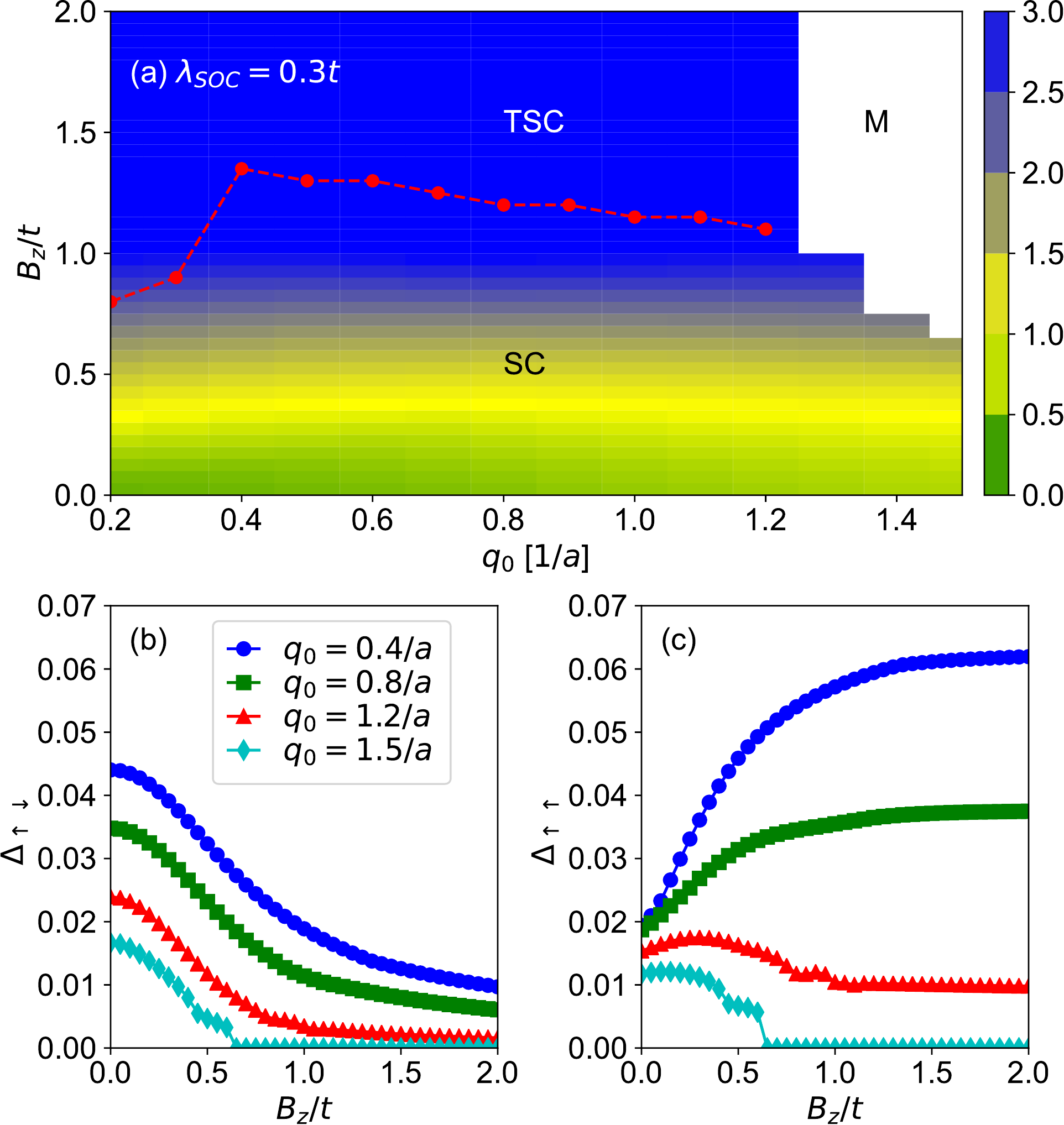}
\caption{\label{Fig:fig1} 
(a) The phase diagram in the plane of the magnetic field $B_z$ and the scattering range $q_0$. The $z$-axis represents the ratio $R$ between spin-up triplet and singlet pairing fields. Here, SC, TSC, and M denote normal superconducting, topological superconducting, and metallic states, respectively. (b) The evolution of the singlet pairing field $\Delta_{\uparrow\downarrow}$ with the magnetic field $B_z$. (c) The evolution of the spin-up triplet pairing field $\Delta_{\uparrow\uparrow}$ with the magnetic field. Here, we set $\lambda_\text{SOC}=0.3t$.}
\end{figure}

{\it Superconductivity.} 
The superconductivity with the interfacial electron-phonon interaction is studied under an external out-of-plane magnetic field. Figure~\ref{Fig:fig1}(a) shows the $q_0$-$B_z$ phase diagram at $\lambda_\text{SOC}=0.3t$. 
A phase transition from a trivial superconducting (SC) state to the topological superconducting (TSC) state occurs as the magnetic field increases when $q_0<1.2/a$. The red dashed curve represents the phase boundary between these two phases. The critical magnetic field $B_c$ for this phase transition exhibits a nonmonotonic behavior. Here, $B_c$ rapidly increases with $q_0$ for $q_0<0.4/a$ and then slightly decreases. The TSC state is replaced by a metallic (M) state when $q_0>1.2/a$, indicating the spoilage of superconductivity. For a Holstein model with a uniform {\eph} coupling, the external magnetic field rapidly suppresses the SC state (see the supplemental material~\cite{supple}). 
These results imply that a short-ranged forward {\eph} interaction stabilizes superconductivity against an applied magnetic field, a rather crucial fact to enable field-induced topology in a superconductor.

\begin{figure}[t]
\center\includegraphics[width=\columnwidth]{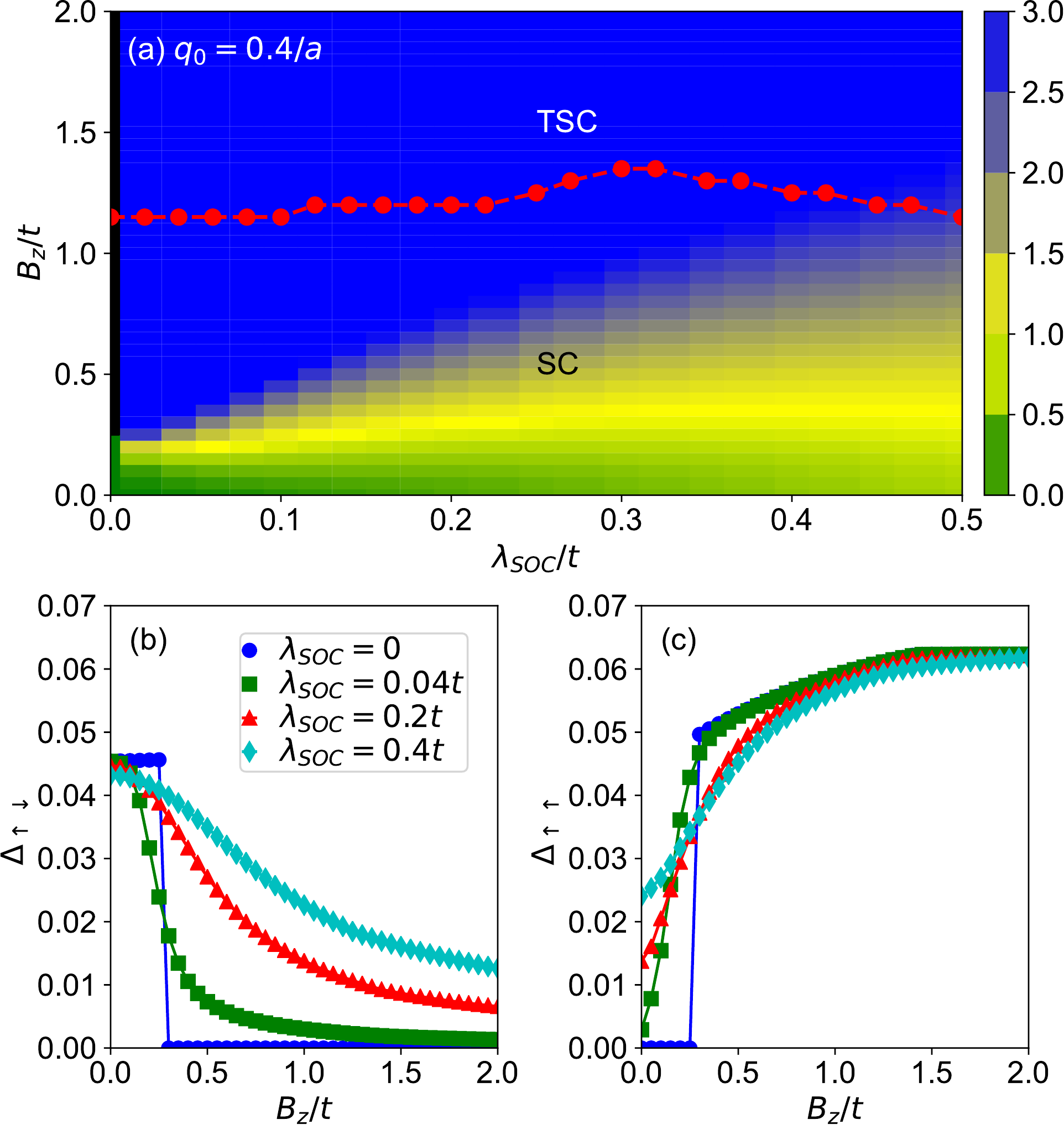}
\caption{\label{Fig:fig2} 
(a) The phase diagram in the plane of the magnetic field $B_z$ and the spin-orbit coupling $\lambda_\text{SOC}$. The $z$-axis represents the ratio $R$ between spin-up triplet and singlet pairing fields. (b) The evolution of the singlet pairing field $\Delta_{\uparrow\downarrow}$ with the magnetic field $B_z$. (c) The evolution of the spin-up triplet pairing field $\Delta_{\uparrow\uparrow}$ with the magnetic field $B_z$. Here, we set $q_0=0.4/a$, where $a$ is the lattice constant.
}
\end{figure}

The persistence of the superconductivity under a magnetic field at small $q_0$ is attributed to the emergence of a triplet pairing potential. In the following, we will demonstrate this argument using a mean-field approach. In the second-order approximation, the two-particle interaction mediated by the {\eph} interaction is given by  $V=\sum_{{\bf k},{\bf k}^\prime,{\bf q},\sigma,\sigma^\prime}U_{\bf q} c^{\dagger}_{{\bf k}+{\bf q},\sigma}c^{\phantom\dagger}_{{\bf k},\sigma}c^{\dagger}_{{\bf k}^\prime-{\bf q},\sigma^\prime}c^{\phantom\dagger}_{{\bf k}^\prime,\sigma^\prime}$. This interaction is represented by the Feynman diagram shown in Fig.~\ref{Fig:fig0}(b). In the Bardeen–Cooper–Schrieffer (BCS) theory~\cite{BCS1957}, this interaction is decoupled into pairing terms, and a pair of electrons with momenta $\bf k$ and $-\bf k$ and spin $\sigma$ and $\sigma^\prime$ is coupled via an effective pairing potential, given by $V_{{\bf k},\sigma,\sigma^\prime}=\sum_{\bf q}U_{\bf q}\langle c_{{\bf k}+{\bf q},\sigma}^{\dagger}c_{-{\bf k}-{\bf q},\sigma^\prime}^{\dagger}\rangle$. When $U_{\bf q}$ is independent of $\bf q$, the effective potential becomes $V_{{\bf k},\sigma,\sigma^\prime}=U\sum_{{\bf k}^\prime}\langle c_{{\bf k}^\prime,\sigma}^{\dagger}c_{-{\bf k}^\prime,\sigma^\prime}^{\dagger}\rangle$. In this case, the effective potential on two electrons with the same spin is zero because $\langle c_{{\bf k}^\prime,\sigma}^{\dagger}c_{-{\bf k}^\prime,\sigma}^{\dagger}\rangle$ has to be odd in parity. Therefore, a uniform {\eph} interaction only supports the $s$-wave singlet pairing. However, when $U_{\bf q}$ is nonzero only at small momentum, say $|{\bf q}|=0$, the effective potential on two electrons with the same spin becomes $U\langle c_{{\bf k},\sigma}^{\dagger}c_{-{\bf k},\sigma}^{\dagger}\rangle$, which is sketched in Fig.~\ref{Fig:fig0}(c). In the presence of the spin-orbit coupling or a magnetic field, the effective triplet pairing potential is no longer zero. This nonzero triplet pairing potential can further enhance the triplet pairing field under a magnetic field. 

\begin{figure*}[t]
\center\includegraphics[width=\textwidth]{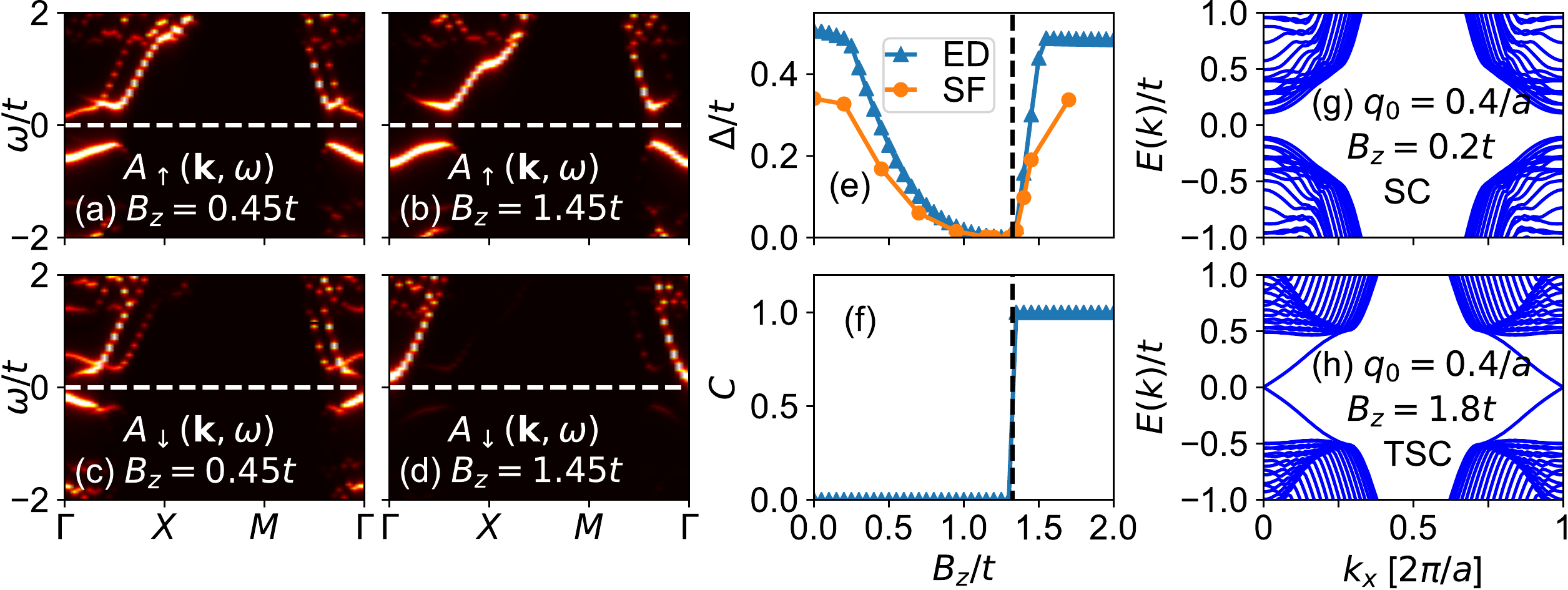}
\caption{\label{Fig:fig3} Topological phase transition. 
Panels (a) and (b) plot the spectral functions $A_{\uparrow}({\bf k},\omega)$ for the spin-up electrons at $B_z=0.45t$ and $1.45t$, respectively. Panels (c) and (d) plot the corresponding spectral functions $A_{\downarrow}({\bf k},\omega)$ for the spin-down electrons. (e) The evolution of the Bogoliubov gap $\Delta$ as a function of the magnetic field. (f) The Chern number $C$ as a function of the magnetic field $B_z$.
Panels (g) and (h) plot quasiparticle band structures of a finite thickness ribbon in the Normal superconducting (SC) and topological superconducting states (TSC), respectively.
Here, the RSOC is set as  $\lambda_\text{SOC}=0.3t$ and the scattering range is $q_0=0.4/a$.}
\end{figure*}

To shed light on the {\eph} coupling induced pairing symmetry, we examine the evolution of both spin-singlet and triplet pairing fields under a magnetic field. 
The pairing field is a momentum dependent quantity. Here, we focus on the average value of the pairing field and define the singlet $\Delta_{\uparrow\downarrow}$ , spin-up triplet $\Delta_{\uparrow\uparrow}$, and spin-down triplet pairing $\Delta_{\downarrow\downarrow}$ fields as $\text{mean}(|\langle c_{{\bf k},\uparrow}c_{-{\bf k},\downarrow}\rangle|)$, $\text{mean}(|\langle c_{{\bf k},\uparrow}c_{-{\bf k},\uparrow}\rangle|)$, and $\text{mean}(|\langle c_{{\bf k},\downarrow}c_{-{\bf k},\downarrow}\rangle|)$, respectively.
Figures~\ref{Fig:fig1}(b) and~\ref{Fig:fig1}(c) plot $\Delta_{\uparrow\downarrow}$ and $\Delta_{\uparrow\uparrow}$, respectively, for different values of $q_0$.
The {\eph} interaction induces the $s$-wave singlet SC state in the absence of the magnetic field and RSOC~\cite{Shaozhinpj}. RSOC hybridizes spin-up and spin-down electrons, leading to a nonzero triplet pairing field at $B_z=0$. Our nonzero $\Delta_{\uparrow\uparrow}$ at $B_z=0$ shown in Fig.~\ref{Fig:fig1}(c) is consistent with this prediction.
Figure~\ref{Fig:fig1}(b) shows that the singlet pairing field is continuously suppressed by the magnetic field; however, $\Delta_{\uparrow\downarrow}$ does not vanish under a strong magnetic field when the scattering range $q_0$ is small. This nonzero $\Delta_{\uparrow\downarrow}$ under a strong magnetic field is attributed to the existence of the nonzero triplet pairing field and the hybridization between spin-up and spin-down electrons. In contrast to the singlet pairing field, the spin-up triplet pairing field is enhanced by the magnetic field when the scattering range is small, consistent with our previous mean-field analysis. To gain insight into the evolution of the triplet pairing field under the magnetic field, the color map of Fig.~\ref{Fig:fig1}(a) shows the ratio $R$ between $\Delta_{\uparrow\uparrow}$ and $\Delta_{\uparrow\downarrow}$. We observe that the pairing field is dominated by the singlet (spin-up triplet) field as $B_z<0.4$ ($B_z>0.5$). Due to the nonzero $\lambda_\text{SOC}$, the TSC state shown in Fig.~\ref{Fig:fig1}(a) is not a pure triplet pairing state.

Now we analyze the effect of the RSOC. Figure~\ref{Fig:fig2}(a) plots the $B_z$-$\lambda_\text{SOC}$ phase diagram at $q_0=0.4/a$, and Figs.~\ref{Fig:fig2}(b) and~\ref{Fig:fig2}(c) show the corresponding singlet and spin-up triplet pairing fields as a function of the magnetic field, respectively. At $\lambda_\text{SOC}=0$, the phase transition from the SC state to the TSC state occurs at $B_z=1.15t$ when the band of spin-down electrons moves above the Fermi surface. The critical magnetic field weakly depends on $\lambda_\text{SOC}$. Although $\lambda_\text{SOC}$ has a tiny effect on the critical magnetic field, $\lambda_\text{SOC}$ stabilizes the singlet pairing field under a strong magnetic field, making $\Delta_{\uparrow\downarrow}$ decrease slowly. This result can be directly observed in Fig.~\ref{Fig:fig2}(b). Besides, we observe a sudden change in $\Delta_{\uparrow\downarrow}$ and $\Delta_{\uparrow\uparrow}$ at $B_z=0.3t$ and $\lambda_\text{SOC}=0$, indicating that a pure singlet SC state becomes a pure triplet SC state. This transition is not accompanied by the topological phase transition, implying that the TSC phase transition is not induced by the change of the symmetry of the pairing field.

{\it Topological phase transition.} Our topological phase transition is induced by a Lifshitz transition. To see this, we plot the momentum-dependent spectral function $A_{\sigma}({\bf k},\omega)$ in Figs.~\ref{Fig:fig3}(a)-~\ref{Fig:fig3}(d) with $q_0=0.4/a$ and $\lambda_\text{SOC}=0.3t$. $A_{\sigma}({\bf k},\omega)$ is obtained from the analytic continuation with the P\'{a}de approximation. 
Here, the band gap near the Fermi level shown in Figs.~\ref{Fig:fig3} (a)-~\ref{Fig:fig3} (d) is not the Bogoliubov gap because the analytic continuation is performed on the normal Green's function. 
The spin-orbit coupling breaks the spin SU(2) symmetry and generates two bands at non-high-symmetry momenta. 
We label a band with a smaller (larger) Fermi momentum as band $\alpha$ ($\eta$). 
At $B_z=0$,
spin-up and spin-down electrons have an equivalent weight on both bands. At a fixed electron density, the magnetic field transfers spin-up (spin-down) electrons to the $\eta$ ($\alpha$) band. With increasing the magnetic field, the $\alpha$ band at the $\Gamma$ point slowly moves from below to above the Fermi surface.
At $B_z=1.3t$, the $\alpha$ band goes across the Fermi surface at the $\Gamma$ point  (see supplemental material~\cite{supple}) and is dominated by the spin-down electrons in the low-energy region. The ground state becomes the TSC state when band $\alpha$ moves above the Fermi surface ($B_z>1.3t$).

To see the detail of the topological phase transition, 
Fig.~\ref{Fig:fig3}(e) plots the Bogoliubov gap $\Delta$ obtained from the spectral function (SF) $A_\sigma({\bf k},\omega)$ 
and the exact diagonalization (ED) of the effective single-particle Hamiltonian $H'$. $\Delta$ from the spectral function is obtained by searching for the minimal excitation energy. With increasing magnetic field $B_z$, both Bogoliubov gaps gradually approach zero at $B_z=1.3t$ and rapidly increase by further increasing $B_z$, indicating that a topological phase transition occurs at $B_z=1.3t$. 
Figure~\ref{Fig:fig3}(f) plots the Chern number $C$ as a function of $B_z$. $C$ discontinuously changes from 0 to 1 at $B_z=1.3t$, 
confirming that the Lifshitz transition and the topological transition coincide.


The TSC state at high fields is further confirmed by examining the edge spectrum. 
For this purpose, 
we construct an effective single-particle Hamiltonian on a finite thickness ribbon 
with a periodic boundary condition along the $x$-direction and an open boundary condition along the $y$-direction. In this case, the self-energy $\Sigma(k_x, r_y)$ is obtained from the Fourier transform of $\Sigma({\bf k})$. We cut off $\Sigma(k_x,r_y)$ at a distance $|r_y|=3a$ because $\Sigma(k_x,r_y)$ is negligible when $|r_y|>3a$ (see details in the supplemental material~\cite{supple}). Figures~\ref{Fig:fig3}(g) and~\ref{Fig:fig3}(h) plot the ribbon-geometry spectrum with $\lambda_\text{SOC}=0.3t$ and a ribbon thickness of $100$. We observe a nonzero Bogoliubov gap with $q_0=0.4/a$ and $B_z=0.2t$ and gapless chiral Majorana edge modes with $q_0=0.4/a$ and $B_z=1.8t$. This result confirms the TSC state in the latter.

{\it Conclusions - } We have uncovered that the forward {\eph} interaction is in favor of generating spin-triplet superconductivity under a magnetic field. 
We also predict a magnetic-field-induced transition ($\lambda_{\text{SOC}}=0$) or crossover ($\lambda_{\text{SOC}}\neq0$) from a singlet-dominant to a triplet-dominant pairing phase.
These results allow us to design a new Rashba-free protocol to build artificial chiral topological superconductor with the help of a ``phonon-proximity" effect. Since the triplet superconductivity is intrinsically generated in the add-on layer, we do expect the superconducting gap to be generically larger than that of a superconducting-proxitimized system, thus providing a new pathway towards high-temperature non-Abelian Majorana platforms~\cite{KITAEV20032,NayakRMP2008,sarma2015,BiaoPNAS2018}.

The relaxation of Rashba interaction greatly facilitates the materialization of our platform. For example, the 2D electron gas can be obtained by using the metal-oxide-semiconductor field-effect transistors~\cite{KlitzingPRL1980} or the ZnO/ZnMgO heterostructure. Since the electron gas can feature a rather high mobility, the disorder effect will be suppressed to guarantee the robustness of Majorana physics in the topological superconductor. In addition,  $\text{TiO}_2$-terminated $\text{SrTiO}_3$ is a good candidate for phonon substrate, because it is expected to feature a short-ranged forward phonon scattering ($q_0 \approx 0.1/a-0.3/a$) to the two-dimensional electron gas due to the vibration of oxygen atoms~\cite{Lee_2015}. We hope our work will inspire more theoretical and experimental efforts to explore Majorana physics from an electron-phonon origin. 

{\it Acknowledgements.} S.L. and S.O. are supported by the U.S. Department of Energy, Office of Science, National Quantum Information Science Research Centers, Quantum Science
Center. L.-H.H. and R.-X. Z. are supported by a startup fund from University of Tennessee.
This research used resources of the Compute and Data Environment for Science (CADES) at the Oak Ridge National Laboratory, which is supported by the Office of Science of the U.S. Department of Energy under Contract No. DE-AC05-00OR22725.  

\bibliography{main}

\begin{thebibliography}{60}%
\makeatletter
\providecommand \@ifxundefined [1]{%
 \@ifx{#1\undefined}
}%
\providecommand \@ifnum [1]{%
 \ifnum #1\expandafter \@firstoftwo
 \else \expandafter \@secondoftwo
 \fi
}%
\providecommand \@ifx [1]{%
 \ifx #1\expandafter \@firstoftwo
 \else \expandafter \@secondoftwo
 \fi
}%
\providecommand \natexlab [1]{#1}%
\providecommand \enquote  [1]{``#1''}%
\providecommand \bibnamefont  [1]{#1}%
\providecommand \bibfnamefont [1]{#1}%
\providecommand \citenamefont [1]{#1}%
\providecommand \href@noop [0]{\@secondoftwo}%
\providecommand \href [0]{\begingroup \@sanitize@url \@href}%
\providecommand \@href[1]{\@@startlink{#1}\@@href}%
\providecommand \@@href[1]{\endgroup#1\@@endlink}%
\providecommand \@sanitize@url [0]{\catcode `\\12\catcode `\$12\catcode
  `\&12\catcode `\#12\catcode `\^12\catcode `\_12\catcode `\%12\relax}%
\providecommand \@@startlink[1]{}%
\providecommand \@@endlink[0]{}%
\providecommand \url  [0]{\begingroup\@sanitize@url \@url }%
\providecommand \@url [1]{\endgroup\@href {#1}{\urlprefix }}%
\providecommand \urlprefix  [0]{URL }%
\providecommand \Eprint [0]{\href }%
\providecommand \doibase [0]{http://dx.doi.org/}%
\providecommand \selectlanguage [0]{\@gobble}%
\providecommand \bibinfo  [0]{\@secondoftwo}%
\providecommand \bibfield  [0]{\@secondoftwo}%
\providecommand \translation [1]{[#1]}%
\providecommand \BibitemOpen [0]{}%
\providecommand \bibitemStop [0]{}%
\providecommand \bibitemNoStop [0]{.\EOS\space}%
\providecommand \EOS [0]{\spacefactor3000\relax}%
\providecommand \BibitemShut  [1]{\csname bibitem#1\endcsname}%
\let\auto@bib@innerbib\@empty
\bibitem [{\citenamefont {Freedman}\ \emph {et~al.}(2006)\citenamefont
  {Freedman}, \citenamefont {Nayak},\ and\ \citenamefont
  {Walker}}]{FreedmanPRB2006}%
  \BibitemOpen
  \bibfield  {author} {\bibinfo {author} {\bibfnamefont {M.}~\bibnamefont
  {Freedman}}, \bibinfo {author} {\bibfnamefont {C.}~\bibnamefont {Nayak}}, \
  and\ \bibinfo {author} {\bibfnamefont {K.}~\bibnamefont {Walker}},\ }\href
  {\doibase 10.1103/PhysRevB.73.245307} {\bibfield  {journal} {\bibinfo
  {journal} {Phys. Rev. B}\ }\textbf {\bibinfo {volume} {73}},\ \bibinfo
  {pages} {245307} (\bibinfo {year} {2006})}\BibitemShut {NoStop}%
\bibitem [{\citenamefont {Nayak}\ \emph
  {et~al.}(2008{\natexlab{a}})\citenamefont {Nayak}, \citenamefont {Simon},
  \citenamefont {Stern}, \citenamefont {Freedman},\ and\ \citenamefont
  {Das~Sarma}}]{Nayak2008}%
  \BibitemOpen
  \bibfield  {author} {\bibinfo {author} {\bibfnamefont {C.}~\bibnamefont
  {Nayak}}, \bibinfo {author} {\bibfnamefont {S.~H.}\ \bibnamefont {Simon}},
  \bibinfo {author} {\bibfnamefont {A.}~\bibnamefont {Stern}}, \bibinfo
  {author} {\bibfnamefont {M.}~\bibnamefont {Freedman}}, \ and\ \bibinfo
  {author} {\bibfnamefont {S.}~\bibnamefont {Das~Sarma}},\ }\href {\doibase
  10.1103/RevModPhys.80.1083} {\bibfield  {journal} {\bibinfo  {journal} {Rev.
  Mod. Phys.}\ }\textbf {\bibinfo {volume} {80}},\ \bibinfo {pages} {1083}
  (\bibinfo {year} {2008}{\natexlab{a}})}\BibitemShut {NoStop}%
\bibitem [{\citenamefont {Field}\ and\ \citenamefont
  {Simula}(2018)}]{Field_2018}%
  \BibitemOpen
  \bibfield  {author} {\bibinfo {author} {\bibfnamefont {B.}~\bibnamefont
  {Field}}\ and\ \bibinfo {author} {\bibfnamefont {T.}~\bibnamefont {Simula}},\
  }\href {\doibase 10.1088/2058-9565/aacad2} {\bibfield  {journal} {\bibinfo
  {journal} {Quantum Science and Technology}\ }\textbf {\bibinfo {volume}
  {3}},\ \bibinfo {pages} {045004} (\bibinfo {year} {2018})}\BibitemShut
  {NoStop}%
\bibitem [{\citenamefont {Lian}\ \emph {et~al.}(2018)\citenamefont {Lian},
  \citenamefont {Sun}, \citenamefont {Vaezi}, \citenamefont {Qi},\ and\
  \citenamefont {Zhang}}]{BiaoPNAS2018}%
  \BibitemOpen
  \bibfield  {author} {\bibinfo {author} {\bibfnamefont {B.}~\bibnamefont
  {Lian}}, \bibinfo {author} {\bibfnamefont {X.-Q.}\ \bibnamefont {Sun}},
  \bibinfo {author} {\bibfnamefont {A.}~\bibnamefont {Vaezi}}, \bibinfo
  {author} {\bibfnamefont {X.-L.}\ \bibnamefont {Qi}}, \ and\ \bibinfo {author}
  {\bibfnamefont {S.-C.}\ \bibnamefont {Zhang}},\ }\href {\doibase
  10.1073/pnas.1810003115} {\bibfield  {journal} {\bibinfo  {journal}
  {Proceedings of the National Academy of Sciences}\ }\textbf {\bibinfo
  {volume} {115}},\ \bibinfo {pages} {10938} (\bibinfo {year}
  {2018})}\BibitemShut {NoStop}%
\bibitem [{\citenamefont {Lutchyn}\ \emph {et~al.}(2018)\citenamefont
  {Lutchyn}, \citenamefont {Bakkers}, \citenamefont {Kouwenhoven},
  \citenamefont {Krogstrup}, \citenamefont {Marcus},\ and\ \citenamefont
  {Oreg}}]{Lutchyn}%
  \BibitemOpen
  \bibfield  {author} {\bibinfo {author} {\bibfnamefont {R.~M.}\ \bibnamefont
  {Lutchyn}}, \bibinfo {author} {\bibfnamefont {E.~P. A.~M.}\ \bibnamefont
  {Bakkers}}, \bibinfo {author} {\bibfnamefont {L.~P.}\ \bibnamefont
  {Kouwenhoven}}, \bibinfo {author} {\bibfnamefont {P.}~\bibnamefont
  {Krogstrup}}, \bibinfo {author} {\bibfnamefont {C.~M.}\ \bibnamefont
  {Marcus}}, \ and\ \bibinfo {author} {\bibfnamefont {Y.}~\bibnamefont
  {Oreg}},\ }\href {\doibase 10.1038/s41578-018-0003-1} {\bibfield  {journal}
  {\bibinfo  {journal} {Nature Reviews Materials}\ }\textbf {\bibinfo {volume}
  {3}},\ \bibinfo {pages} {52} (\bibinfo {year} {2018})}\BibitemShut {NoStop}%
\bibitem [{\citenamefont {McMillan}(1968)}]{McM1968}%
  \BibitemOpen
  \bibfield  {author} {\bibinfo {author} {\bibfnamefont {W.~L.}\ \bibnamefont
  {McMillan}},\ }\href {\doibase 10.1103/PhysRev.175.537} {\bibfield  {journal}
  {\bibinfo  {journal} {Phys. Rev.}\ }\textbf {\bibinfo {volume} {175}},\
  \bibinfo {pages} {537} (\bibinfo {year} {1968})}\BibitemShut {NoStop}%
\bibitem [{\citenamefont {Fu}\ and\ \citenamefont {Kane}(2008)}]{KanePRL2008}%
  \BibitemOpen
  \bibfield  {author} {\bibinfo {author} {\bibfnamefont {L.}~\bibnamefont
  {Fu}}\ and\ \bibinfo {author} {\bibfnamefont {C.~L.}\ \bibnamefont {Kane}},\
  }\href {\doibase 10.1103/PhysRevLett.100.096407} {\bibfield  {journal}
  {\bibinfo  {journal} {Phys. Rev. Lett.}\ }\textbf {\bibinfo {volume} {100}},\
  \bibinfo {pages} {096407} (\bibinfo {year} {2008})}\BibitemShut {NoStop}%
\bibitem [{\citenamefont {Qi}\ \emph {et~al.}(2009)\citenamefont {Qi},
  \citenamefont {Hughes}, \citenamefont {Raghu},\ and\ \citenamefont
  {Zhang}}]{QiPRL2009}%
  \BibitemOpen
  \bibfield  {author} {\bibinfo {author} {\bibfnamefont {X.-L.}\ \bibnamefont
  {Qi}}, \bibinfo {author} {\bibfnamefont {T.~L.}\ \bibnamefont {Hughes}},
  \bibinfo {author} {\bibfnamefont {S.}~\bibnamefont {Raghu}}, \ and\ \bibinfo
  {author} {\bibfnamefont {S.-C.}\ \bibnamefont {Zhang}},\ }\href {\doibase
  10.1103/PhysRevLett.102.187001} {\bibfield  {journal} {\bibinfo  {journal}
  {Phys. Rev. Lett.}\ }\textbf {\bibinfo {volume} {102}},\ \bibinfo {pages}
  {187001} (\bibinfo {year} {2009})}\BibitemShut {NoStop}%
\bibitem [{\citenamefont {Alicea}(2012)}]{Alicea_2012}%
  \BibitemOpen
  \bibfield  {author} {\bibinfo {author} {\bibfnamefont {J.}~\bibnamefont
  {Alicea}},\ }\href {\doibase 10.1088/0034-4885/75/7/076501} {\bibfield
  {journal} {\bibinfo  {journal} {Reports on Progress in Physics}\ }\textbf
  {\bibinfo {volume} {75}},\ \bibinfo {pages} {076501} (\bibinfo {year}
  {2012})}\BibitemShut {NoStop}%
\bibitem [{\citenamefont {Beenakker}(2013)}]{Beenakker2013}%
  \BibitemOpen
  \bibfield  {author} {\bibinfo {author} {\bibfnamefont {C.}~\bibnamefont
  {Beenakker}},\ }\href {\doibase 10.1146/annurev-conmatphys-030212-184337}
  {\bibfield  {journal} {\bibinfo  {journal} {Annual Review of Condensed Matter
  Physics}\ }\textbf {\bibinfo {volume} {4}},\ \bibinfo {pages} {113} (\bibinfo
  {year} {2013})},\ \Eprint
  {http://arxiv.org/abs/https://doi.org/10.1146/annurev-conmatphys-030212-184337}
  {https://doi.org/10.1146/annurev-conmatphys-030212-184337} \BibitemShut
  {NoStop}%
\bibitem [{\citenamefont {Jiang}\ and\ \citenamefont {Wu}(2013)}]{Jiang_2013}%
  \BibitemOpen
  \bibfield  {author} {\bibinfo {author} {\bibfnamefont {J.-H.}\ \bibnamefont
  {Jiang}}\ and\ \bibinfo {author} {\bibfnamefont {S.}~\bibnamefont {Wu}},\
  }\href {\doibase 10.1088/0953-8984/25/5/055701} {\bibfield  {journal}
  {\bibinfo  {journal} {Journal of Physics: Condensed Matter}\ }\textbf
  {\bibinfo {volume} {25}},\ \bibinfo {pages} {055701} (\bibinfo {year}
  {2013})}\BibitemShut {NoStop}%
\bibitem [{\citenamefont {Xu}\ \emph {et~al.}(2014)\citenamefont {Xu},
  \citenamefont {Liu}, \citenamefont {Wang}, \citenamefont {Ge}, \citenamefont
  {Liu}, \citenamefont {Yang}, \citenamefont {Chen}, \citenamefont {Liu},
  \citenamefont {Xu}, \citenamefont {Gao}, \citenamefont {Qian}, \citenamefont
  {Zhang},\ and\ \citenamefont {Jia}}]{XuPRL2014}%
  \BibitemOpen
  \bibfield  {author} {\bibinfo {author} {\bibfnamefont {J.-P.}\ \bibnamefont
  {Xu}}, \bibinfo {author} {\bibfnamefont {C.}~\bibnamefont {Liu}}, \bibinfo
  {author} {\bibfnamefont {M.-X.}\ \bibnamefont {Wang}}, \bibinfo {author}
  {\bibfnamefont {J.}~\bibnamefont {Ge}}, \bibinfo {author} {\bibfnamefont
  {Z.-L.}\ \bibnamefont {Liu}}, \bibinfo {author} {\bibfnamefont
  {X.}~\bibnamefont {Yang}}, \bibinfo {author} {\bibfnamefont {Y.}~\bibnamefont
  {Chen}}, \bibinfo {author} {\bibfnamefont {Y.}~\bibnamefont {Liu}}, \bibinfo
  {author} {\bibfnamefont {Z.-A.}\ \bibnamefont {Xu}}, \bibinfo {author}
  {\bibfnamefont {C.-L.}\ \bibnamefont {Gao}}, \bibinfo {author} {\bibfnamefont
  {D.}~\bibnamefont {Qian}}, \bibinfo {author} {\bibfnamefont {F.-C.}\
  \bibnamefont {Zhang}}, \ and\ \bibinfo {author} {\bibfnamefont {J.-F.}\
  \bibnamefont {Jia}},\ }\href {\doibase 10.1103/PhysRevLett.112.217001}
  {\bibfield  {journal} {\bibinfo  {journal} {Phys. Rev. Lett.}\ }\textbf
  {\bibinfo {volume} {112}},\ \bibinfo {pages} {217001} (\bibinfo {year}
  {2014})}\BibitemShut {NoStop}%
\bibitem [{\citenamefont {Sajadi}\ \emph {et~al.}(2018)\citenamefont {Sajadi},
  \citenamefont {Palomaki}, \citenamefont {Fei}, \citenamefont {Zhao},
  \citenamefont {Bement}, \citenamefont {Olsen}, \citenamefont {Luescher},
  \citenamefont {Xu}, \citenamefont {Folk},\ and\ \citenamefont
  {Cobden}}]{Sajadi2018}%
  \BibitemOpen
  \bibfield  {author} {\bibinfo {author} {\bibfnamefont {E.}~\bibnamefont
  {Sajadi}}, \bibinfo {author} {\bibfnamefont {T.}~\bibnamefont {Palomaki}},
  \bibinfo {author} {\bibfnamefont {Z.}~\bibnamefont {Fei}}, \bibinfo {author}
  {\bibfnamefont {W.}~\bibnamefont {Zhao}}, \bibinfo {author} {\bibfnamefont
  {P.}~\bibnamefont {Bement}}, \bibinfo {author} {\bibfnamefont
  {C.}~\bibnamefont {Olsen}}, \bibinfo {author} {\bibfnamefont
  {S.}~\bibnamefont {Luescher}}, \bibinfo {author} {\bibfnamefont
  {X.}~\bibnamefont {Xu}}, \bibinfo {author} {\bibfnamefont {J.~A.}\
  \bibnamefont {Folk}}, \ and\ \bibinfo {author} {\bibfnamefont {D.~H.}\
  \bibnamefont {Cobden}},\ }\href {\doibase 10.1126/science.aar4426} {\bibfield
   {journal} {\bibinfo  {journal} {Science}\ }\textbf {\bibinfo {volume}
  {362}},\ \bibinfo {pages} {922} (\bibinfo {year} {2018})},\ \Eprint
  {http://arxiv.org/abs/https://www.science.org/doi/pdf/10.1126/science.aar4426}
  {https://www.science.org/doi/pdf/10.1126/science.aar4426} \BibitemShut
  {NoStop}%
\bibitem [{\citenamefont {Zhang}\ \emph {et~al.}(2018)\citenamefont {Zhang},
  \citenamefont {Yaji}, \citenamefont {Hashimoto}, \citenamefont {Ota},
  \citenamefont {Kondo}, \citenamefont {Okazaki}, \citenamefont {Wang},
  \citenamefont {Wen}, \citenamefont {Gu}, \citenamefont {Ding},\ and\
  \citenamefont {Shin}}]{ZhangScience2018}%
  \BibitemOpen
  \bibfield  {author} {\bibinfo {author} {\bibfnamefont {P.}~\bibnamefont
  {Zhang}}, \bibinfo {author} {\bibfnamefont {K.}~\bibnamefont {Yaji}},
  \bibinfo {author} {\bibfnamefont {T.}~\bibnamefont {Hashimoto}}, \bibinfo
  {author} {\bibfnamefont {Y.}~\bibnamefont {Ota}}, \bibinfo {author}
  {\bibfnamefont {T.}~\bibnamefont {Kondo}}, \bibinfo {author} {\bibfnamefont
  {K.}~\bibnamefont {Okazaki}}, \bibinfo {author} {\bibfnamefont
  {Z.}~\bibnamefont {Wang}}, \bibinfo {author} {\bibfnamefont {J.}~\bibnamefont
  {Wen}}, \bibinfo {author} {\bibfnamefont {G.~D.}\ \bibnamefont {Gu}},
  \bibinfo {author} {\bibfnamefont {H.}~\bibnamefont {Ding}}, \ and\ \bibinfo
  {author} {\bibfnamefont {S.}~\bibnamefont {Shin}},\ }\href {\doibase
  10.1126/science.aan4596} {\bibfield  {journal} {\bibinfo  {journal}
  {Science}\ }\textbf {\bibinfo {volume} {360}},\ \bibinfo {pages} {182}
  (\bibinfo {year} {2018})}\BibitemShut {NoStop}%
\bibitem [{\citenamefont {Wang}\ \emph {et~al.}(2018)\citenamefont {Wang},
  \citenamefont {Lingyuan}, \citenamefont {Fan}, \citenamefont {Chen},
  \citenamefont {Zhu}, \citenamefont {Liu}, \citenamefont {Cao}, \citenamefont
  {Sun}, \citenamefont {Du}, \citenamefont {Schineeloch}, \citenamefont
  {Zhong}, \citenamefont {Gu}, \citenamefont {Fu}, \citenamefont {Ding},\ and\
  \citenamefont {Gao}}]{WangScience2018}%
  \BibitemOpen
  \bibfield  {author} {\bibinfo {author} {\bibfnamefont {D.}~\bibnamefont
  {Wang}}, \bibinfo {author} {\bibfnamefont {K.}~\bibnamefont {Lingyuan}},
  \bibinfo {author} {\bibfnamefont {P.}~\bibnamefont {Fan}}, \bibinfo {author}
  {\bibfnamefont {H.}~\bibnamefont {Chen}}, \bibinfo {author} {\bibfnamefont
  {S.}~\bibnamefont {Zhu}}, \bibinfo {author} {\bibfnamefont {W.}~\bibnamefont
  {Liu}}, \bibinfo {author} {\bibfnamefont {L.}~\bibnamefont {Cao}}, \bibinfo
  {author} {\bibfnamefont {Y.}~\bibnamefont {Sun}}, \bibinfo {author}
  {\bibfnamefont {S.}~\bibnamefont {Du}}, \bibinfo {author} {\bibfnamefont
  {J.}~\bibnamefont {Schineeloch}}, \bibinfo {author} {\bibfnamefont
  {R.}~\bibnamefont {Zhong}}, \bibinfo {author} {\bibfnamefont
  {G.}~\bibnamefont {Gu}}, \bibinfo {author} {\bibfnamefont {L.}~\bibnamefont
  {Fu}}, \bibinfo {author} {\bibfnamefont {H.}~\bibnamefont {Ding}}, \ and\
  \bibinfo {author} {\bibfnamefont {H.-j.}\ \bibnamefont {Gao}},\ }\href
  {\doibase 10.1126/science.aao1797} {\bibfield  {journal} {\bibinfo  {journal}
  {Science}\ }\textbf {\bibinfo {volume} {362}},\ \bibinfo {pages} {333}
  (\bibinfo {year} {2018})}\BibitemShut {NoStop}%
\bibitem [{\citenamefont {Lee}\ \emph {et~al.}(2019)\citenamefont {Lee},
  \citenamefont {Hazra}, \citenamefont {Randeria},\ and\ \citenamefont
  {Trivedi}}]{Lee2019PRB}%
  \BibitemOpen
  \bibfield  {author} {\bibinfo {author} {\bibfnamefont {K.}~\bibnamefont
  {Lee}}, \bibinfo {author} {\bibfnamefont {T.}~\bibnamefont {Hazra}}, \bibinfo
  {author} {\bibfnamefont {M.}~\bibnamefont {Randeria}}, \ and\ \bibinfo
  {author} {\bibfnamefont {N.}~\bibnamefont {Trivedi}},\ }\href {\doibase
  10.1103/PhysRevB.99.184514} {\bibfield  {journal} {\bibinfo  {journal} {Phys.
  Rev. B}\ }\textbf {\bibinfo {volume} {99}},\ \bibinfo {pages} {184514}
  (\bibinfo {year} {2019})}\BibitemShut {NoStop}%
\bibitem [{\citenamefont {Machida}\ \emph {et~al.}(2019)\citenamefont
  {Machida}, \citenamefont {Sun}, \citenamefont {Pyon}, \citenamefont {Takeda},
  \citenamefont {Kohsaka}, \citenamefont {Hanaguri}, \citenamefont {Sasagawa},\
  and\ \citenamefont {Tamegai}}]{Machida2019}%
  \BibitemOpen
  \bibfield  {author} {\bibinfo {author} {\bibfnamefont {T.}~\bibnamefont
  {Machida}}, \bibinfo {author} {\bibfnamefont {Y.}~\bibnamefont {Sun}},
  \bibinfo {author} {\bibfnamefont {S.}~\bibnamefont {Pyon}}, \bibinfo {author}
  {\bibfnamefont {S.}~\bibnamefont {Takeda}}, \bibinfo {author} {\bibfnamefont
  {Y.}~\bibnamefont {Kohsaka}}, \bibinfo {author} {\bibfnamefont
  {T.}~\bibnamefont {Hanaguri}}, \bibinfo {author} {\bibfnamefont
  {T.}~\bibnamefont {Sasagawa}}, \ and\ \bibinfo {author} {\bibfnamefont
  {T.}~\bibnamefont {Tamegai}},\ }\href {\doibase 10.1038/s41563-019-0397-1}
  {\bibfield  {journal} {\bibinfo  {journal} {Nature Materials}\ }\textbf
  {\bibinfo {volume} {18}},\ \bibinfo {pages} {811} (\bibinfo {year}
  {2019})}\BibitemShut {NoStop}%
\bibitem [{\citenamefont {Zhu}\ \emph {et~al.}(2019)\citenamefont {Zhu},
  \citenamefont {Kong}, \citenamefont {Cao}, \citenamefont {Chen},
  \citenamefont {Papaj}, \citenamefont {Du}, \citenamefont {Xing},
  \citenamefont {Liu}, \citenamefont {Wang}, \citenamefont {Shen},
  \citenamefont {Yang}, \citenamefont {Schneeloch}, \citenamefont {Zhong},
  \citenamefont {Gu}, \citenamefont {Fu}, \citenamefont {Zhang}, \citenamefont
  {Ding},\ and\ \citenamefont {Gao}}]{Zhuscience2019}%
  \BibitemOpen
  \bibfield  {author} {\bibinfo {author} {\bibfnamefont {S.}~\bibnamefont
  {Zhu}}, \bibinfo {author} {\bibfnamefont {L.}~\bibnamefont {Kong}}, \bibinfo
  {author} {\bibfnamefont {L.}~\bibnamefont {Cao}}, \bibinfo {author}
  {\bibfnamefont {H.}~\bibnamefont {Chen}}, \bibinfo {author} {\bibfnamefont
  {M.}~\bibnamefont {Papaj}}, \bibinfo {author} {\bibfnamefont
  {S.}~\bibnamefont {Du}}, \bibinfo {author} {\bibfnamefont {Y.}~\bibnamefont
  {Xing}}, \bibinfo {author} {\bibfnamefont {W.}~\bibnamefont {Liu}}, \bibinfo
  {author} {\bibfnamefont {D.}~\bibnamefont {Wang}}, \bibinfo {author}
  {\bibfnamefont {C.}~\bibnamefont {Shen}}, \bibinfo {author} {\bibfnamefont
  {F.}~\bibnamefont {Yang}}, \bibinfo {author} {\bibfnamefont {J.}~\bibnamefont
  {Schneeloch}}, \bibinfo {author} {\bibfnamefont {R.}~\bibnamefont {Zhong}},
  \bibinfo {author} {\bibfnamefont {G.}~\bibnamefont {Gu}}, \bibinfo {author}
  {\bibfnamefont {L.}~\bibnamefont {Fu}}, \bibinfo {author} {\bibfnamefont
  {Y.-y.}\ \bibnamefont {Zhang}}, \bibinfo {author} {\bibfnamefont
  {H.}~\bibnamefont {Ding}}, \ and\ \bibinfo {author} {\bibfnamefont {H.-j.}\
  \bibnamefont {Gao}},\ }\href {\doibase 10.1126/science.aax0274} {\bibfield
  {journal} {\bibinfo  {journal} {Science}\ }\textbf {\bibinfo {volume}
  {367}},\ \bibinfo {pages} {189} (\bibinfo {year} {2019})}\BibitemShut
  {NoStop}%
\bibitem [{\citenamefont {Rameau}\ \emph {et~al.}(2019)\citenamefont {Rameau},
  \citenamefont {Zaki}, \citenamefont {Gu}, \citenamefont {Johnson},\ and\
  \citenamefont {Weinert}}]{RameauPRB2019}%
  \BibitemOpen
  \bibfield  {author} {\bibinfo {author} {\bibfnamefont {J.~D.}\ \bibnamefont
  {Rameau}}, \bibinfo {author} {\bibfnamefont {N.}~\bibnamefont {Zaki}},
  \bibinfo {author} {\bibfnamefont {G.~D.}\ \bibnamefont {Gu}}, \bibinfo
  {author} {\bibfnamefont {P.~D.}\ \bibnamefont {Johnson}}, \ and\ \bibinfo
  {author} {\bibfnamefont {M.}~\bibnamefont {Weinert}},\ }\href {\doibase
  10.1103/PhysRevB.99.205117} {\bibfield  {journal} {\bibinfo  {journal} {Phys.
  Rev. B}\ }\textbf {\bibinfo {volume} {99}},\ \bibinfo {pages} {205117}
  (\bibinfo {year} {2019})}\BibitemShut {NoStop}%
\bibitem [{\citenamefont {Trang}\ \emph {et~al.}(2020)\citenamefont {Trang},
  \citenamefont {Shimamura}, \citenamefont {Nakayama}, \citenamefont {Souma},
  \citenamefont {Sugawara}, \citenamefont {Watanabe}, \citenamefont {Yamauchi},
  \citenamefont {Oguchi}, \citenamefont {Segawa}, \citenamefont {Takahashi},
  \citenamefont {Ando},\ and\ \citenamefont {Sato}}]{Trang2020}%
  \BibitemOpen
  \bibfield  {author} {\bibinfo {author} {\bibfnamefont {C.~X.}\ \bibnamefont
  {Trang}}, \bibinfo {author} {\bibfnamefont {N.}~\bibnamefont {Shimamura}},
  \bibinfo {author} {\bibfnamefont {K.}~\bibnamefont {Nakayama}}, \bibinfo
  {author} {\bibfnamefont {S.}~\bibnamefont {Souma}}, \bibinfo {author}
  {\bibfnamefont {K.}~\bibnamefont {Sugawara}}, \bibinfo {author}
  {\bibfnamefont {I.}~\bibnamefont {Watanabe}}, \bibinfo {author}
  {\bibfnamefont {K.}~\bibnamefont {Yamauchi}}, \bibinfo {author}
  {\bibfnamefont {T.}~\bibnamefont {Oguchi}}, \bibinfo {author} {\bibfnamefont
  {K.}~\bibnamefont {Segawa}}, \bibinfo {author} {\bibfnamefont
  {T.}~\bibnamefont {Takahashi}}, \bibinfo {author} {\bibfnamefont
  {Y.}~\bibnamefont {Ando}}, \ and\ \bibinfo {author} {\bibfnamefont
  {T.}~\bibnamefont {Sato}},\ }\href {\doibase 10.1038/s41467-019-13946-0}
  {\bibfield  {journal} {\bibinfo  {journal} {Nature Communications}\ }\textbf
  {\bibinfo {volume} {11}},\ \bibinfo {pages} {159} (\bibinfo {year}
  {2020})}\BibitemShut {NoStop}%
\bibitem [{\citenamefont {Wang}\ \emph {et~al.}(2020)\citenamefont {Wang},
  \citenamefont {Rodriguez}, \citenamefont {Jiao}, \citenamefont {Howard},
  \citenamefont {Graham}, \citenamefont {Gu}, \citenamefont {Hughes},
  \citenamefont {Morr},\ and\ \citenamefont {Madhavan}}]{Wangscience2020}%
  \BibitemOpen
  \bibfield  {author} {\bibinfo {author} {\bibfnamefont {Z.}~\bibnamefont
  {Wang}}, \bibinfo {author} {\bibfnamefont {J.~O.}\ \bibnamefont {Rodriguez}},
  \bibinfo {author} {\bibfnamefont {L.}~\bibnamefont {Jiao}}, \bibinfo {author}
  {\bibfnamefont {S.}~\bibnamefont {Howard}}, \bibinfo {author} {\bibfnamefont
  {M.}~\bibnamefont {Graham}}, \bibinfo {author} {\bibfnamefont {G.~D.}\
  \bibnamefont {Gu}}, \bibinfo {author} {\bibfnamefont {T.~L.}\ \bibnamefont
  {Hughes}}, \bibinfo {author} {\bibfnamefont {D.~K.}\ \bibnamefont {Morr}}, \
  and\ \bibinfo {author} {\bibfnamefont {V.}~\bibnamefont {Madhavan}},\ }\href
  {\doibase 10.1126/science.aaw8419} {\bibfield  {journal} {\bibinfo  {journal}
  {Science}\ }\textbf {\bibinfo {volume} {367}},\ \bibinfo {pages} {104}
  (\bibinfo {year} {2020})}\BibitemShut {NoStop}%
\bibitem [{\citenamefont {Chen}\ \emph {et~al.}(2020)\citenamefont {Chen},
  \citenamefont {Jiang}, \citenamefont {Zhang}, \citenamefont {Liu},
  \citenamefont {Liu}, \citenamefont {Wang},\ and\ \citenamefont
  {Wang}}]{ChengNP2020}%
  \BibitemOpen
  \bibfield  {author} {\bibinfo {author} {\bibfnamefont {C.}~\bibnamefont
  {Chen}}, \bibinfo {author} {\bibfnamefont {K.}~\bibnamefont {Jiang}},
  \bibinfo {author} {\bibfnamefont {Y.}~\bibnamefont {Zhang}}, \bibinfo
  {author} {\bibfnamefont {C.}~\bibnamefont {Liu}}, \bibinfo {author}
  {\bibfnamefont {Y.}~\bibnamefont {Liu}}, \bibinfo {author} {\bibfnamefont
  {Z.}~\bibnamefont {Wang}}, \ and\ \bibinfo {author} {\bibfnamefont
  {J.}~\bibnamefont {Wang}},\ }\href {\doibase 10.1038/s41567-020-0813-0}
  {\bibfield  {journal} {\bibinfo  {journal} {Nature Physics}\ }\textbf
  {\bibinfo {volume} {16}},\ \bibinfo {pages} {536} (\bibinfo {year}
  {2020})}\BibitemShut {NoStop}%
\bibitem [{\citenamefont {Mascot}\ \emph {et~al.}(2021)\citenamefont {Mascot},
  \citenamefont {Cocklin}, \citenamefont {Graham}, \citenamefont {Mashkoori},
  \citenamefont {Rachel},\ and\ \citenamefont {Morr}}]{Mascot2021}%
  \BibitemOpen
  \bibfield  {author} {\bibinfo {author} {\bibfnamefont {E.}~\bibnamefont
  {Mascot}}, \bibinfo {author} {\bibfnamefont {S.}~\bibnamefont {Cocklin}},
  \bibinfo {author} {\bibfnamefont {M.}~\bibnamefont {Graham}}, \bibinfo
  {author} {\bibfnamefont {M.}~\bibnamefont {Mashkoori}}, \bibinfo {author}
  {\bibfnamefont {S.}~\bibnamefont {Rachel}}, \ and\ \bibinfo {author}
  {\bibfnamefont {D.~K.}\ \bibnamefont {Morr}},\ }\href@noop {} {\bibfield
  {journal} {\bibinfo  {journal} {arXiv: 2102.05116}\ } (\bibinfo {year}
  {2021})}\BibitemShut {NoStop}%
\bibitem [{\citenamefont {Zhang}\ and\ \citenamefont
  {Das~Sarma}(2021)}]{ZhangPRL2021}%
  \BibitemOpen
  \bibfield  {author} {\bibinfo {author} {\bibfnamefont {R.-X.}\ \bibnamefont
  {Zhang}}\ and\ \bibinfo {author} {\bibfnamefont {S.}~\bibnamefont
  {Das~Sarma}},\ }\href {\doibase 10.1103/PhysRevLett.126.137001} {\bibfield
  {journal} {\bibinfo  {journal} {Phys. Rev. Lett.}\ }\textbf {\bibinfo
  {volume} {126}},\ \bibinfo {pages} {137001} (\bibinfo {year}
  {2021})}\BibitemShut {NoStop}%
\bibitem [{\citenamefont {Qi}\ and\ \citenamefont
  {Zhang}(2011)}]{QiRevModPhys}%
  \BibitemOpen
  \bibfield  {author} {\bibinfo {author} {\bibfnamefont {X.-L.}\ \bibnamefont
  {Qi}}\ and\ \bibinfo {author} {\bibfnamefont {S.-C.}\ \bibnamefont {Zhang}},\
  }\href {\doibase 10.1103/RevModPhys.83.1057} {\bibfield  {journal} {\bibinfo
  {journal} {Rev. Mod. Phys.}\ }\textbf {\bibinfo {volume} {83}},\ \bibinfo
  {pages} {1057} (\bibinfo {year} {2011})}\BibitemShut {NoStop}%
\bibitem [{\citenamefont {Yang}\ \emph {et~al.}(2021)\citenamefont {Yang},
  \citenamefont {Luo}, \citenamefont {Yi}, \citenamefont {Shi}, \citenamefont
  {Zhou},\ and\ \citenamefont {qing Zheng}}]{Jirscience2021}%
  \BibitemOpen
  \bibfield  {author} {\bibinfo {author} {\bibfnamefont {J.}~\bibnamefont
  {Yang}}, \bibinfo {author} {\bibfnamefont {J.}~\bibnamefont {Luo}}, \bibinfo
  {author} {\bibfnamefont {C.}~\bibnamefont {Yi}}, \bibinfo {author}
  {\bibfnamefont {Y.}~\bibnamefont {Shi}}, \bibinfo {author} {\bibfnamefont
  {Y.}~\bibnamefont {Zhou}}, \ and\ \bibinfo {author} {\bibfnamefont
  {G.}~\bibnamefont {qing Zheng}},\ }\href {\doibase 10.1126/sciadv.abl4432}
  {\bibfield  {journal} {\bibinfo  {journal} {Science Advances}\ }\textbf
  {\bibinfo {volume} {7}},\ \bibinfo {pages} {eabl4432} (\bibinfo {year}
  {2021})},\ \Eprint
  {http://arxiv.org/abs/https://www.science.org/doi/pdf/10.1126/sciadv.abl4432}
  {https://www.science.org/doi/pdf/10.1126/sciadv.abl4432} \BibitemShut
  {NoStop}%
\bibitem [{\citenamefont {Chou}\ \emph {et~al.}(2021)\citenamefont {Chou},
  \citenamefont {Wu}, \citenamefont {Sau},\ and\ \citenamefont
  {Das~Sarma}}]{ChouPRL2021}%
  \BibitemOpen
  \bibfield  {author} {\bibinfo {author} {\bibfnamefont {Y.-Z.}\ \bibnamefont
  {Chou}}, \bibinfo {author} {\bibfnamefont {F.}~\bibnamefont {Wu}}, \bibinfo
  {author} {\bibfnamefont {J.~D.}\ \bibnamefont {Sau}}, \ and\ \bibinfo
  {author} {\bibfnamefont {S.}~\bibnamefont {Das~Sarma}},\ }\href {\doibase
  10.1103/PhysRevLett.127.217001} {\bibfield  {journal} {\bibinfo  {journal}
  {Phys. Rev. Lett.}\ }\textbf {\bibinfo {volume} {127}},\ \bibinfo {pages}
  {217001} (\bibinfo {year} {2021})}\BibitemShut {NoStop}%
\bibitem [{\citenamefont {Hor}\ \emph {et~al.}(2010)\citenamefont {Hor},
  \citenamefont {Williams}, \citenamefont {Checkelsky}, \citenamefont
  {Roushan}, \citenamefont {Seo}, \citenamefont {Xu}, \citenamefont
  {Zandbergen}, \citenamefont {Yazdani}, \citenamefont {Ong},\ and\
  \citenamefont {Cava}}]{HorPRL2010}%
  \BibitemOpen
  \bibfield  {author} {\bibinfo {author} {\bibfnamefont {Y.~S.}\ \bibnamefont
  {Hor}}, \bibinfo {author} {\bibfnamefont {A.~J.}\ \bibnamefont {Williams}},
  \bibinfo {author} {\bibfnamefont {J.~G.}\ \bibnamefont {Checkelsky}},
  \bibinfo {author} {\bibfnamefont {P.}~\bibnamefont {Roushan}}, \bibinfo
  {author} {\bibfnamefont {J.}~\bibnamefont {Seo}}, \bibinfo {author}
  {\bibfnamefont {Q.}~\bibnamefont {Xu}}, \bibinfo {author} {\bibfnamefont
  {H.~W.}\ \bibnamefont {Zandbergen}}, \bibinfo {author} {\bibfnamefont
  {A.}~\bibnamefont {Yazdani}}, \bibinfo {author} {\bibfnamefont {N.~P.}\
  \bibnamefont {Ong}}, \ and\ \bibinfo {author} {\bibfnamefont {R.~J.}\
  \bibnamefont {Cava}},\ }\href {\doibase 10.1103/PhysRevLett.104.057001}
  {\bibfield  {journal} {\bibinfo  {journal} {Phys. Rev. Lett.}\ }\textbf
  {\bibinfo {volume} {104}},\ \bibinfo {pages} {057001} (\bibinfo {year}
  {2010})}\BibitemShut {NoStop}%
\bibitem [{\citenamefont {Fu}\ and\ \citenamefont {Berg}(2010)}]{LiangPRL2010}%
  \BibitemOpen
  \bibfield  {author} {\bibinfo {author} {\bibfnamefont {L.}~\bibnamefont
  {Fu}}\ and\ \bibinfo {author} {\bibfnamefont {E.}~\bibnamefont {Berg}},\
  }\href {\doibase 10.1103/PhysRevLett.105.097001} {\bibfield  {journal}
  {\bibinfo  {journal} {Phys. Rev. Lett.}\ }\textbf {\bibinfo {volume} {105}},\
  \bibinfo {pages} {097001} (\bibinfo {year} {2010})}\BibitemShut {NoStop}%
\bibitem [{\citenamefont {Mourik}\ \emph {et~al.}(2012)\citenamefont {Mourik},
  \citenamefont {Zuo}, \citenamefont {Frolov}, \citenamefont {Plissard},
  \citenamefont {Bakkers},\ and\ \citenamefont {Kouwenhoven}}]{Mourik2012}%
  \BibitemOpen
  \bibfield  {author} {\bibinfo {author} {\bibfnamefont {V.}~\bibnamefont
  {Mourik}}, \bibinfo {author} {\bibfnamefont {K.}~\bibnamefont {Zuo}},
  \bibinfo {author} {\bibfnamefont {S.~M.}\ \bibnamefont {Frolov}}, \bibinfo
  {author} {\bibfnamefont {S.~R.}\ \bibnamefont {Plissard}}, \bibinfo {author}
  {\bibfnamefont {E.~P. A.~M.}\ \bibnamefont {Bakkers}}, \ and\ \bibinfo
  {author} {\bibfnamefont {L.~P.}\ \bibnamefont {Kouwenhoven}},\ }\href
  {\doibase 10.1126/science.1222360} {\bibfield  {journal} {\bibinfo  {journal}
  {Science}\ }\textbf {\bibinfo {volume} {336}},\ \bibinfo {pages} {1003}
  (\bibinfo {year} {2012})},\ \Eprint
  {http://arxiv.org/abs/https://www.science.org/doi/pdf/10.1126/science.1222360}
  {https://www.science.org/doi/pdf/10.1126/science.1222360} \BibitemShut
  {NoStop}%
\bibitem [{\citenamefont {Kallin}(2012)}]{Kallin_2012}%
  \BibitemOpen
  \bibfield  {author} {\bibinfo {author} {\bibfnamefont {C.}~\bibnamefont
  {Kallin}},\ }\href {\doibase 10.1088/0034-4885/75/4/042501} {\bibfield
  {journal} {\bibinfo  {journal} {Reports on Progress in Physics}\ }\textbf
  {\bibinfo {volume} {75}},\ \bibinfo {pages} {042501} (\bibinfo {year}
  {2012})}\BibitemShut {NoStop}%
\bibitem [{\citenamefont {Das}\ \emph {et~al.}(2012)\citenamefont {Das},
  \citenamefont {Ronen}, \citenamefont {Most}, \citenamefont {Oreg},
  \citenamefont {Heiblum},\ and\ \citenamefont {Shtrikman}}]{Das2012}%
  \BibitemOpen
  \bibfield  {author} {\bibinfo {author} {\bibfnamefont {A.}~\bibnamefont
  {Das}}, \bibinfo {author} {\bibfnamefont {Y.}~\bibnamefont {Ronen}}, \bibinfo
  {author} {\bibfnamefont {Y.}~\bibnamefont {Most}}, \bibinfo {author}
  {\bibfnamefont {Y.}~\bibnamefont {Oreg}}, \bibinfo {author} {\bibfnamefont
  {M.}~\bibnamefont {Heiblum}}, \ and\ \bibinfo {author} {\bibfnamefont
  {H.}~\bibnamefont {Shtrikman}},\ }\href {\doibase 10.1038/nphys2479}
  {\bibfield  {journal} {\bibinfo  {journal} {Nature Physics}\ }\textbf
  {\bibinfo {volume} {8}},\ \bibinfo {pages} {887} (\bibinfo {year}
  {2012})}\BibitemShut {NoStop}%
\bibitem [{\citenamefont {Xu}\ \emph {et~al.}(2016)\citenamefont {Xu},
  \citenamefont {Lian}, \citenamefont {Tang}, \citenamefont {Qi},\ and\
  \citenamefont {Zhang}}]{GangPRL2016}%
  \BibitemOpen
  \bibfield  {author} {\bibinfo {author} {\bibfnamefont {G.}~\bibnamefont
  {Xu}}, \bibinfo {author} {\bibfnamefont {B.}~\bibnamefont {Lian}}, \bibinfo
  {author} {\bibfnamefont {P.}~\bibnamefont {Tang}}, \bibinfo {author}
  {\bibfnamefont {X.-L.}\ \bibnamefont {Qi}}, \ and\ \bibinfo {author}
  {\bibfnamefont {S.-C.}\ \bibnamefont {Zhang}},\ }\href {\doibase
  10.1103/PhysRevLett.117.047001} {\bibfield  {journal} {\bibinfo  {journal}
  {Phys. Rev. Lett.}\ }\textbf {\bibinfo {volume} {117}},\ \bibinfo {pages}
  {047001} (\bibinfo {year} {2016})}\BibitemShut {NoStop}%
\bibitem [{\citenamefont {Liu}\ \emph {et~al.}(2017)\citenamefont {Liu},
  \citenamefont {Zhang}, \citenamefont {Rice},\ and\ \citenamefont
  {Wang}}]{Liunpj2017}%
  \BibitemOpen
  \bibfield  {author} {\bibinfo {author} {\bibfnamefont {Y.-C.}\ \bibnamefont
  {Liu}}, \bibinfo {author} {\bibfnamefont {F.-C.}\ \bibnamefont {Zhang}},
  \bibinfo {author} {\bibfnamefont {T.~M.}\ \bibnamefont {Rice}}, \ and\
  \bibinfo {author} {\bibfnamefont {Q.-H.}\ \bibnamefont {Wang}},\ }\href
  {\doibase 10.1038/s41535-017-0014-y} {\bibfield  {journal} {\bibinfo
  {journal} {npj Quantum Materials}\ }\textbf {\bibinfo {volume} {12}},\
  \bibinfo {pages} {12} (\bibinfo {year} {2017})}\BibitemShut {NoStop}%
\bibitem [{\citenamefont {Li}\ \emph {et~al.}(2021)\citenamefont {Li},
  \citenamefont {Zheng}, \citenamefont {Fang}, \citenamefont {Zhang},
  \citenamefont {Chen}, \citenamefont {Chen}, \citenamefont {Liang},
  \citenamefont {Shi}, \citenamefont {Pei}, \citenamefont {Xu}, \citenamefont
  {Liu}, \citenamefont {Pan}, \citenamefont {Lu}, \citenamefont {Hashimoto},
  \citenamefont {Barinov}, \citenamefont {Jung}, \citenamefont {Cacho},
  \citenamefont {Wang}, \citenamefont {He}, \citenamefont {Fu}, \citenamefont
  {Zhang}, \citenamefont {Huang}, \citenamefont {Yang}, \citenamefont {Liu},\
  and\ \citenamefont {Chen}}]{LiNC2021}%
  \BibitemOpen
  \bibfield  {author} {\bibinfo {author} {\bibfnamefont {Y.~W.}\ \bibnamefont
  {Li}}, \bibinfo {author} {\bibfnamefont {H.~J.}\ \bibnamefont {Zheng}},
  \bibinfo {author} {\bibfnamefont {Y.~Q.}\ \bibnamefont {Fang}}, \bibinfo
  {author} {\bibfnamefont {D.~Q.}\ \bibnamefont {Zhang}}, \bibinfo {author}
  {\bibfnamefont {Y.~J.}\ \bibnamefont {Chen}}, \bibinfo {author}
  {\bibfnamefont {C.}~\bibnamefont {Chen}}, \bibinfo {author} {\bibfnamefont
  {A.~J.}\ \bibnamefont {Liang}}, \bibinfo {author} {\bibfnamefont {W.~J.}\
  \bibnamefont {Shi}}, \bibinfo {author} {\bibfnamefont {D.}~\bibnamefont
  {Pei}}, \bibinfo {author} {\bibfnamefont {L.~X.}\ \bibnamefont {Xu}},
  \bibinfo {author} {\bibfnamefont {S.}~\bibnamefont {Liu}}, \bibinfo {author}
  {\bibfnamefont {J.}~\bibnamefont {Pan}}, \bibinfo {author} {\bibfnamefont
  {D.~H.}\ \bibnamefont {Lu}}, \bibinfo {author} {\bibfnamefont
  {M.}~\bibnamefont {Hashimoto}}, \bibinfo {author} {\bibfnamefont
  {A.}~\bibnamefont {Barinov}}, \bibinfo {author} {\bibfnamefont {S.~W.}\
  \bibnamefont {Jung}}, \bibinfo {author} {\bibfnamefont {C.}~\bibnamefont
  {Cacho}}, \bibinfo {author} {\bibfnamefont {M.~X.}\ \bibnamefont {Wang}},
  \bibinfo {author} {\bibfnamefont {Y.}~\bibnamefont {He}}, \bibinfo {author}
  {\bibfnamefont {L.}~\bibnamefont {Fu}}, \bibinfo {author} {\bibfnamefont
  {H.~J.}\ \bibnamefont {Zhang}}, \bibinfo {author} {\bibfnamefont {F.~Q.}\
  \bibnamefont {Huang}}, \bibinfo {author} {\bibfnamefont {L.~X.}\ \bibnamefont
  {Yang}}, \bibinfo {author} {\bibfnamefont {Z.~K.}\ \bibnamefont {Liu}}, \
  and\ \bibinfo {author} {\bibfnamefont {Y.~L.}\ \bibnamefont {Chen}},\ }\href
  {\doibase 10.1038/s41467-021-23076-1} {\bibfield  {journal} {\bibinfo
  {journal} {Nature Communication}\ }\textbf {\bibinfo {volume} {12}},\
  \bibinfo {pages} {2874} (\bibinfo {year} {2021})}\BibitemShut {NoStop}%
\bibitem [{\citenamefont {Kezilebieke}\ \emph {et~al.}(2020)\citenamefont
  {Kezilebieke}, \citenamefont {Huda}, \citenamefont {Vaňo}, \citenamefont
  {Aapro}, \citenamefont {Ganguli}, \citenamefont {Silveira}, \citenamefont
  {Głodzik}, \citenamefont {Foster}, \citenamefont {Ojanen},\ and\
  \citenamefont {Liljeroth}}]{KezilebiekeNature}%
  \BibitemOpen
  \bibfield  {author} {\bibinfo {author} {\bibfnamefont {S.}~\bibnamefont
  {Kezilebieke}}, \bibinfo {author} {\bibfnamefont {M.~N.}\ \bibnamefont
  {Huda}}, \bibinfo {author} {\bibfnamefont {V.}~\bibnamefont {Vaňo}},
  \bibinfo {author} {\bibfnamefont {M.}~\bibnamefont {Aapro}}, \bibinfo
  {author} {\bibfnamefont {S.~C.}\ \bibnamefont {Ganguli}}, \bibinfo {author}
  {\bibfnamefont {O.~J.}\ \bibnamefont {Silveira}}, \bibinfo {author}
  {\bibfnamefont {S.}~\bibnamefont {Głodzik}}, \bibinfo {author}
  {\bibfnamefont {A.~S.}\ \bibnamefont {Foster}}, \bibinfo {author}
  {\bibfnamefont {T.}~\bibnamefont {Ojanen}}, \ and\ \bibinfo {author}
  {\bibfnamefont {P.}~\bibnamefont {Liljeroth}},\ }\href {\doibase
  10.1038/s41586-020-2989-y} {\bibfield  {journal} {\bibinfo  {journal}
  {Nature}\ ,\ \bibinfo {pages} {424}} (\bibinfo {year} {2020})}\BibitemShut
  {NoStop}%
\bibitem [{\citenamefont {Aghaee}\ \emph {et~al.}(2022)\citenamefont {Aghaee},
  \citenamefont {Akkala}, \citenamefont {Alam}, \citenamefont {Ali},
  \citenamefont {Ramirez}, \citenamefont {Andrzejczuk}, \citenamefont
  {Antipov}, \citenamefont {Astafev}, \citenamefont {Bauer}, \citenamefont
  {Becker}, \citenamefont {Boddapati}, \citenamefont {Boekhout}, \citenamefont
  {Bommer}, \citenamefont {Hansen}, \citenamefont {Bosma}, \citenamefont
  {Bourdet}, \citenamefont {Boutin}, \citenamefont {Caroff}, \citenamefont
  {Casparis}, \citenamefont {Cassidy}, \citenamefont {Christensen},
  \citenamefont {Clay}, \citenamefont {Cole}, \citenamefont {Corsetti},
  \citenamefont {Cui}, \citenamefont {Dalampiras}, \citenamefont {Dokania},
  \citenamefont {de~Lange}, \citenamefont {de~Moor}, \citenamefont {Saldaña},
  \citenamefont {Fallahi}, \citenamefont {Fathabad}, \citenamefont {Gamble},
  \citenamefont {Gardner}, \citenamefont {Govender}, \citenamefont {Griggio},
  \citenamefont {Grigoryan}, \citenamefont {Gronin}, \citenamefont
  {Gukelberger}, \citenamefont {Heedt}, \citenamefont {Zamorano}, \citenamefont
  {Ho}, \citenamefont {Holgaard}, \citenamefont {Nielsen}, \citenamefont
  {Ingerslev}, \citenamefont {Krogstrup}, \citenamefont {Johansson},
  \citenamefont {Jones}, \citenamefont {Kallaher}, \citenamefont {Karimi},
  \citenamefont {Karzig}, \citenamefont {King}, \citenamefont {Kloster},
  \citenamefont {Knapp}, \citenamefont {Kocon}, \citenamefont {Koski},
  \citenamefont {Kostamo}, \citenamefont {Kumar}, \citenamefont {Laeven},
  \citenamefont {Larsen}, \citenamefont {Li}, \citenamefont {Lindemann},
  \citenamefont {Love}, \citenamefont {Lutchyn}, \citenamefont {Manfra},
  \citenamefont {Memisevic}, \citenamefont {Nayak}, \citenamefont {Nijholt},
  \citenamefont {Madsen}, \citenamefont {Markussen}, \citenamefont {Martinez},
  \citenamefont {McNeil}, \citenamefont {Mullally}, \citenamefont {Nielsen},
  \citenamefont {Nurmohamed}, \citenamefont {O'Farrell}, \citenamefont {Otani},
  \citenamefont {Pauka}, \citenamefont {Petersson}, \citenamefont {Petit},
  \citenamefont {Pikulin}, \citenamefont {Preiss}, \citenamefont {Perez},
  \citenamefont {Rasmussen}, \citenamefont {Rajpalke}, \citenamefont
  {Razmadze}, \citenamefont {Reentila}, \citenamefont {Reilly}, \citenamefont
  {Rouse}, \citenamefont {Sadovskyy}, \citenamefont {Sainiemi}, \citenamefont
  {Schreppler}, \citenamefont {Sidorkin}, \citenamefont {Singh}, \citenamefont
  {Singh}, \citenamefont {Sinha}, \citenamefont {Sohr}, \citenamefont {Stek},
  \citenamefont {Suominen}, \citenamefont {Suter}, \citenamefont {Svidenko},
  \citenamefont {Teicher}, \citenamefont {Temuerhan}, \citenamefont
  {Thiyagarajah}, \citenamefont {Tholapi}, \citenamefont {Thomas},
  \citenamefont {Toomey}, \citenamefont {Upadhyay}, \citenamefont {Urban},
  \citenamefont {Vaitiekėnas}, \citenamefont {Hoogdalem}, \citenamefont
  {Viazmitinov}, \citenamefont {Waddy}, \citenamefont {Woerkom}, \citenamefont
  {Vogel}, \citenamefont {Watson}, \citenamefont {Weston}, \citenamefont
  {Winkler}, \citenamefont {Yang}, \citenamefont {Yau}, \citenamefont {Yi},
  \citenamefont {Yucelen}, \citenamefont {Webster}, \citenamefont {Zeisel},\
  and\ \citenamefont {Zhao}}]{Mortezaarxiv}%
  \BibitemOpen
  \bibfield  {author} {\bibinfo {author} {\bibfnamefont {M.}~\bibnamefont
  {Aghaee}}, \bibinfo {author} {\bibfnamefont {A.}~\bibnamefont {Akkala}},
  \bibinfo {author} {\bibfnamefont {Z.}~\bibnamefont {Alam}}, \bibinfo {author}
  {\bibfnamefont {R.}~\bibnamefont {Ali}}, \bibinfo {author} {\bibfnamefont
  {A.~A.}\ \bibnamefont {Ramirez}}, \bibinfo {author} {\bibfnamefont
  {M.}~\bibnamefont {Andrzejczuk}}, \bibinfo {author} {\bibfnamefont {A.~E.}\
  \bibnamefont {Antipov}}, \bibinfo {author} {\bibfnamefont {M.}~\bibnamefont
  {Astafev}}, \bibinfo {author} {\bibfnamefont {B.}~\bibnamefont {Bauer}},
  \bibinfo {author} {\bibfnamefont {J.}~\bibnamefont {Becker}}, \bibinfo
  {author} {\bibfnamefont {S.}~\bibnamefont {Boddapati}}, \bibinfo {author}
  {\bibfnamefont {F.}~\bibnamefont {Boekhout}}, \bibinfo {author}
  {\bibfnamefont {J.}~\bibnamefont {Bommer}}, \bibinfo {author} {\bibfnamefont
  {E.~B.}\ \bibnamefont {Hansen}}, \bibinfo {author} {\bibfnamefont
  {T.}~\bibnamefont {Bosma}}, \bibinfo {author} {\bibfnamefont
  {L.}~\bibnamefont {Bourdet}}, \bibinfo {author} {\bibfnamefont
  {S.}~\bibnamefont {Boutin}}, \bibinfo {author} {\bibfnamefont
  {P.}~\bibnamefont {Caroff}}, \bibinfo {author} {\bibfnamefont
  {L.}~\bibnamefont {Casparis}}, \bibinfo {author} {\bibfnamefont
  {M.}~\bibnamefont {Cassidy}}, \bibinfo {author} {\bibfnamefont {A.~W.}\
  \bibnamefont {Christensen}}, \bibinfo {author} {\bibfnamefont
  {N.}~\bibnamefont {Clay}}, \bibinfo {author} {\bibfnamefont {W.~S.}\
  \bibnamefont {Cole}}, \bibinfo {author} {\bibfnamefont {F.}~\bibnamefont
  {Corsetti}}, \bibinfo {author} {\bibfnamefont {A.}~\bibnamefont {Cui}},
  \bibinfo {author} {\bibfnamefont {P.}~\bibnamefont {Dalampiras}}, \bibinfo
  {author} {\bibfnamefont {A.}~\bibnamefont {Dokania}}, \bibinfo {author}
  {\bibfnamefont {G.}~\bibnamefont {de~Lange}}, \bibinfo {author}
  {\bibfnamefont {M.}~\bibnamefont {de~Moor}}, \bibinfo {author} {\bibfnamefont
  {J.~C.~E.}\ \bibnamefont {Saldaña}}, \bibinfo {author} {\bibfnamefont
  {S.}~\bibnamefont {Fallahi}}, \bibinfo {author} {\bibfnamefont {Z.~H.}\
  \bibnamefont {Fathabad}}, \bibinfo {author} {\bibfnamefont {J.}~\bibnamefont
  {Gamble}}, \bibinfo {author} {\bibfnamefont {G.}~\bibnamefont {Gardner}},
  \bibinfo {author} {\bibfnamefont {D.}~\bibnamefont {Govender}}, \bibinfo
  {author} {\bibfnamefont {F.}~\bibnamefont {Griggio}}, \bibinfo {author}
  {\bibfnamefont {R.}~\bibnamefont {Grigoryan}}, \bibinfo {author}
  {\bibfnamefont {S.}~\bibnamefont {Gronin}}, \bibinfo {author} {\bibfnamefont
  {J.}~\bibnamefont {Gukelberger}}, \bibinfo {author} {\bibfnamefont
  {S.}~\bibnamefont {Heedt}}, \bibinfo {author} {\bibfnamefont {J.~H.}\
  \bibnamefont {Zamorano}}, \bibinfo {author} {\bibfnamefont {S.}~\bibnamefont
  {Ho}}, \bibinfo {author} {\bibfnamefont {U.~L.}\ \bibnamefont {Holgaard}},
  \bibinfo {author} {\bibfnamefont {W.~H.~P.}\ \bibnamefont {Nielsen}},
  \bibinfo {author} {\bibfnamefont {H.}~\bibnamefont {Ingerslev}}, \bibinfo
  {author} {\bibfnamefont {P.~J.}\ \bibnamefont {Krogstrup}}, \bibinfo {author}
  {\bibfnamefont {L.}~\bibnamefont {Johansson}}, \bibinfo {author}
  {\bibfnamefont {J.}~\bibnamefont {Jones}}, \bibinfo {author} {\bibfnamefont
  {R.}~\bibnamefont {Kallaher}}, \bibinfo {author} {\bibfnamefont
  {F.}~\bibnamefont {Karimi}}, \bibinfo {author} {\bibfnamefont
  {T.}~\bibnamefont {Karzig}}, \bibinfo {author} {\bibfnamefont
  {C.}~\bibnamefont {King}}, \bibinfo {author} {\bibfnamefont {M.~E.}\
  \bibnamefont {Kloster}}, \bibinfo {author} {\bibfnamefont {C.}~\bibnamefont
  {Knapp}}, \bibinfo {author} {\bibfnamefont {D.}~\bibnamefont {Kocon}},
  \bibinfo {author} {\bibfnamefont {J.}~\bibnamefont {Koski}}, \bibinfo
  {author} {\bibfnamefont {P.}~\bibnamefont {Kostamo}}, \bibinfo {author}
  {\bibfnamefont {M.}~\bibnamefont {Kumar}}, \bibinfo {author} {\bibfnamefont
  {T.}~\bibnamefont {Laeven}}, \bibinfo {author} {\bibfnamefont
  {T.}~\bibnamefont {Larsen}}, \bibinfo {author} {\bibfnamefont
  {K.}~\bibnamefont {Li}}, \bibinfo {author} {\bibfnamefont {T.}~\bibnamefont
  {Lindemann}}, \bibinfo {author} {\bibfnamefont {J.}~\bibnamefont {Love}},
  \bibinfo {author} {\bibfnamefont {R.}~\bibnamefont {Lutchyn}}, \bibinfo
  {author} {\bibfnamefont {M.}~\bibnamefont {Manfra}}, \bibinfo {author}
  {\bibfnamefont {E.}~\bibnamefont {Memisevic}}, \bibinfo {author}
  {\bibfnamefont {C.}~\bibnamefont {Nayak}}, \bibinfo {author} {\bibfnamefont
  {B.}~\bibnamefont {Nijholt}}, \bibinfo {author} {\bibfnamefont {M.~H.}\
  \bibnamefont {Madsen}}, \bibinfo {author} {\bibfnamefont {S.}~\bibnamefont
  {Markussen}}, \bibinfo {author} {\bibfnamefont {E.}~\bibnamefont {Martinez}},
  \bibinfo {author} {\bibfnamefont {R.}~\bibnamefont {McNeil}}, \bibinfo
  {author} {\bibfnamefont {A.}~\bibnamefont {Mullally}}, \bibinfo {author}
  {\bibfnamefont {J.}~\bibnamefont {Nielsen}}, \bibinfo {author} {\bibfnamefont
  {A.}~\bibnamefont {Nurmohamed}}, \bibinfo {author} {\bibfnamefont
  {E.}~\bibnamefont {O'Farrell}}, \bibinfo {author} {\bibfnamefont
  {K.}~\bibnamefont {Otani}}, \bibinfo {author} {\bibfnamefont
  {S.}~\bibnamefont {Pauka}}, \bibinfo {author} {\bibfnamefont
  {K.}~\bibnamefont {Petersson}}, \bibinfo {author} {\bibfnamefont
  {L.}~\bibnamefont {Petit}}, \bibinfo {author} {\bibfnamefont
  {D.}~\bibnamefont {Pikulin}}, \bibinfo {author} {\bibfnamefont
  {F.}~\bibnamefont {Preiss}}, \bibinfo {author} {\bibfnamefont {M.~Q.}\
  \bibnamefont {Perez}}, \bibinfo {author} {\bibfnamefont {K.}~\bibnamefont
  {Rasmussen}}, \bibinfo {author} {\bibfnamefont {M.}~\bibnamefont {Rajpalke}},
  \bibinfo {author} {\bibfnamefont {D.}~\bibnamefont {Razmadze}}, \bibinfo
  {author} {\bibfnamefont {O.}~\bibnamefont {Reentila}}, \bibinfo {author}
  {\bibfnamefont {D.}~\bibnamefont {Reilly}}, \bibinfo {author} {\bibfnamefont
  {R.}~\bibnamefont {Rouse}}, \bibinfo {author} {\bibfnamefont
  {I.}~\bibnamefont {Sadovskyy}}, \bibinfo {author} {\bibfnamefont
  {L.}~\bibnamefont {Sainiemi}}, \bibinfo {author} {\bibfnamefont
  {S.}~\bibnamefont {Schreppler}}, \bibinfo {author} {\bibfnamefont
  {V.}~\bibnamefont {Sidorkin}}, \bibinfo {author} {\bibfnamefont
  {A.}~\bibnamefont {Singh}}, \bibinfo {author} {\bibfnamefont
  {S.}~\bibnamefont {Singh}}, \bibinfo {author} {\bibfnamefont
  {S.}~\bibnamefont {Sinha}}, \bibinfo {author} {\bibfnamefont
  {P.}~\bibnamefont {Sohr}}, \bibinfo {author} {\bibfnamefont {L.}~\bibnamefont
  {Stek}}, \bibinfo {author} {\bibfnamefont {H.}~\bibnamefont {Suominen}},
  \bibinfo {author} {\bibfnamefont {J.}~\bibnamefont {Suter}}, \bibinfo
  {author} {\bibfnamefont {V.}~\bibnamefont {Svidenko}}, \bibinfo {author}
  {\bibfnamefont {S.}~\bibnamefont {Teicher}}, \bibinfo {author} {\bibfnamefont
  {M.}~\bibnamefont {Temuerhan}}, \bibinfo {author} {\bibfnamefont
  {N.}~\bibnamefont {Thiyagarajah}}, \bibinfo {author} {\bibfnamefont
  {R.}~\bibnamefont {Tholapi}}, \bibinfo {author} {\bibfnamefont
  {M.}~\bibnamefont {Thomas}}, \bibinfo {author} {\bibfnamefont
  {E.}~\bibnamefont {Toomey}}, \bibinfo {author} {\bibfnamefont
  {S.}~\bibnamefont {Upadhyay}}, \bibinfo {author} {\bibfnamefont
  {I.}~\bibnamefont {Urban}}, \bibinfo {author} {\bibfnamefont
  {S.}~\bibnamefont {Vaitiekėnas}}, \bibinfo {author} {\bibfnamefont {K.~V.}\
  \bibnamefont {Hoogdalem}}, \bibinfo {author} {\bibfnamefont {D.~V.}\
  \bibnamefont {Viazmitinov}}, \bibinfo {author} {\bibfnamefont
  {S.}~\bibnamefont {Waddy}}, \bibinfo {author} {\bibfnamefont {D.~V.}\
  \bibnamefont {Woerkom}}, \bibinfo {author} {\bibfnamefont {D.}~\bibnamefont
  {Vogel}}, \bibinfo {author} {\bibfnamefont {J.}~\bibnamefont {Watson}},
  \bibinfo {author} {\bibfnamefont {J.}~\bibnamefont {Weston}}, \bibinfo
  {author} {\bibfnamefont {G.~W.}\ \bibnamefont {Winkler}}, \bibinfo {author}
  {\bibfnamefont {C.~K.}\ \bibnamefont {Yang}}, \bibinfo {author}
  {\bibfnamefont {S.}~\bibnamefont {Yau}}, \bibinfo {author} {\bibfnamefont
  {D.}~\bibnamefont {Yi}}, \bibinfo {author} {\bibfnamefont {E.}~\bibnamefont
  {Yucelen}}, \bibinfo {author} {\bibfnamefont {A.}~\bibnamefont {Webster}},
  \bibinfo {author} {\bibfnamefont {R.}~\bibnamefont {Zeisel}}, \ and\ \bibinfo
  {author} {\bibfnamefont {R.}~\bibnamefont {Zhao}},\ }\href@noop {} {\enquote
  {\bibinfo {title} {Inas-al hybrid devices passing the topological gap
  protocol},}\ } (\bibinfo {year} {2022}),\ \Eprint
  {http://arxiv.org/abs/arXiv:2207.02472} {arXiv:2207.02472} \BibitemShut
  {NoStop}%
\bibitem [{\citenamefont {Hui}\ \emph {et~al.}(2015)\citenamefont {Hui},
  \citenamefont {Sau},\ and\ \citenamefont {Das~Sarma}}]{HuiPRB2015}%
  \BibitemOpen
  \bibfield  {author} {\bibinfo {author} {\bibfnamefont {H.-Y.}\ \bibnamefont
  {Hui}}, \bibinfo {author} {\bibfnamefont {J.~D.}\ \bibnamefont {Sau}}, \ and\
  \bibinfo {author} {\bibfnamefont {S.}~\bibnamefont {Das~Sarma}},\ }\href
  {\doibase 10.1103/PhysRevB.92.174512} {\bibfield  {journal} {\bibinfo
  {journal} {Phys. Rev. B}\ }\textbf {\bibinfo {volume} {92}},\ \bibinfo
  {pages} {174512} (\bibinfo {year} {2015})}\BibitemShut {NoStop}%
\bibitem [{\citenamefont {Lu}\ \emph {et~al.}(2020)\citenamefont {Lu},
  \citenamefont {Virtanen},\ and\ \citenamefont {Heikkil\"a}}]{YaoPRB2020}%
  \BibitemOpen
  \bibfield  {author} {\bibinfo {author} {\bibfnamefont {Y.}~\bibnamefont
  {Lu}}, \bibinfo {author} {\bibfnamefont {P.}~\bibnamefont {Virtanen}}, \ and\
  \bibinfo {author} {\bibfnamefont {T.~T.}\ \bibnamefont {Heikkil\"a}},\ }\href
  {\doibase 10.1103/PhysRevB.102.224510} {\bibfield  {journal} {\bibinfo
  {journal} {Phys. Rev. B}\ }\textbf {\bibinfo {volume} {102}},\ \bibinfo
  {pages} {224510} (\bibinfo {year} {2020})}\BibitemShut {NoStop}%
\bibitem [{\citenamefont {Stanescu}\ and\ \citenamefont
  {Sarma}(2022)}]{Tudor2022}%
  \BibitemOpen
  \bibfield  {author} {\bibinfo {author} {\bibfnamefont {T.~D.}\ \bibnamefont
  {Stanescu}}\ and\ \bibinfo {author} {\bibfnamefont {S.~D.}\ \bibnamefont
  {Sarma}},\ }\href@noop {} {\enquote {\bibinfo {title} {Proximity-induced
  superconductivity generated by thin films: Effects of fermi surface mismatch
  and disorder in the superconductor},}\ } (\bibinfo {year} {2022}),\ \Eprint
  {http://arxiv.org/abs/arXiv:2206.13526} {arXiv:2206.13526} \BibitemShut
  {NoStop}%
\bibitem [{\citenamefont {Lee}\ \emph {et~al.}(2014)\citenamefont {Lee},
  \citenamefont {Schmitt}, \citenamefont {Moore}, \citenamefont {Johnston},
  \citenamefont {Cui}, \citenamefont {Li}, \citenamefont {Yi}, \citenamefont
  {Liu}, \citenamefont {Hashimoto}, \citenamefont {Zhang}, \citenamefont {Lu},
  \citenamefont {Devereaux}, \citenamefont {Lee},\ and\ \citenamefont
  {Shen}}]{LeeNature2014}%
  \BibitemOpen
  \bibfield  {author} {\bibinfo {author} {\bibfnamefont {J.~J.}\ \bibnamefont
  {Lee}}, \bibinfo {author} {\bibfnamefont {F.~T.}\ \bibnamefont {Schmitt}},
  \bibinfo {author} {\bibfnamefont {R.~G.}\ \bibnamefont {Moore}}, \bibinfo
  {author} {\bibfnamefont {S.}~\bibnamefont {Johnston}}, \bibinfo {author}
  {\bibfnamefont {Y.-T.}\ \bibnamefont {Cui}}, \bibinfo {author} {\bibfnamefont
  {W.}~\bibnamefont {Li}}, \bibinfo {author} {\bibfnamefont {M.}~\bibnamefont
  {Yi}}, \bibinfo {author} {\bibfnamefont {Z.~K.}\ \bibnamefont {Liu}},
  \bibinfo {author} {\bibfnamefont {M.}~\bibnamefont {Hashimoto}}, \bibinfo
  {author} {\bibfnamefont {Y.}~\bibnamefont {Zhang}}, \bibinfo {author}
  {\bibfnamefont {D.~H.}\ \bibnamefont {Lu}}, \bibinfo {author} {\bibfnamefont
  {T.~P.}\ \bibnamefont {Devereaux}}, \bibinfo {author} {\bibfnamefont {D.-H.}\
  \bibnamefont {Lee}}, \ and\ \bibinfo {author} {\bibfnamefont {Z.-X.}\
  \bibnamefont {Shen}},\ }\href {\doibase 10.1038/nature13894} {\bibfield
  {journal} {\bibinfo  {journal} {Nature}\ }\textbf {\bibinfo {volume} {515}},\
  \bibinfo {pages} {245} (\bibinfo {year} {2014})}\BibitemShut {NoStop}%
\bibitem [{\citenamefont {Lee}(2015)}]{Lee_2015}%
  \BibitemOpen
  \bibfield  {author} {\bibinfo {author} {\bibfnamefont {D.-H.}\ \bibnamefont
  {Lee}},\ }\href {\doibase 10.1088/1674-1056/24/11/117405} {\bibfield
  {journal} {\bibinfo  {journal} {Chinese Physics B}\ }\textbf {\bibinfo
  {volume} {24}},\ \bibinfo {pages} {117405} (\bibinfo {year}
  {2015})}\BibitemShut {NoStop}%
\bibitem [{\citenamefont {Rademaker}\ \emph {et~al.}(2016)\citenamefont
  {Rademaker}, \citenamefont {Wang}, \citenamefont {Berlijn},\ and\
  \citenamefont {Johnston}}]{Rademaker_2016}%
  \BibitemOpen
  \bibfield  {author} {\bibinfo {author} {\bibfnamefont {L.}~\bibnamefont
  {Rademaker}}, \bibinfo {author} {\bibfnamefont {Y.}~\bibnamefont {Wang}},
  \bibinfo {author} {\bibfnamefont {T.}~\bibnamefont {Berlijn}}, \ and\
  \bibinfo {author} {\bibfnamefont {S.}~\bibnamefont {Johnston}},\ }\href
  {\doibase 10.1088/1367-2630/18/2/022001} {\bibfield  {journal} {\bibinfo
  {journal} {New Journal of Physics}\ }\textbf {\bibinfo {volume} {18}},\
  \bibinfo {pages} {022001} (\bibinfo {year} {2016})}\BibitemShut {NoStop}%
\bibitem [{\citenamefont {Zhang}\ \emph {et~al.}(2016)\citenamefont {Zhang},
  \citenamefont {Guan}, \citenamefont {Jia}, \citenamefont {Liu}, \citenamefont
  {Wang}, \citenamefont {Li}, \citenamefont {Wang}, \citenamefont {Ma},
  \citenamefont {Xue}, \citenamefont {Zhang}, \citenamefont {Plummer},
  \citenamefont {Zhu},\ and\ \citenamefont {Guo}}]{ZhangPRB2016}%
  \BibitemOpen
  \bibfield  {author} {\bibinfo {author} {\bibfnamefont {S.}~\bibnamefont
  {Zhang}}, \bibinfo {author} {\bibfnamefont {J.}~\bibnamefont {Guan}},
  \bibinfo {author} {\bibfnamefont {X.}~\bibnamefont {Jia}}, \bibinfo {author}
  {\bibfnamefont {B.}~\bibnamefont {Liu}}, \bibinfo {author} {\bibfnamefont
  {W.}~\bibnamefont {Wang}}, \bibinfo {author} {\bibfnamefont {F.}~\bibnamefont
  {Li}}, \bibinfo {author} {\bibfnamefont {L.}~\bibnamefont {Wang}}, \bibinfo
  {author} {\bibfnamefont {X.}~\bibnamefont {Ma}}, \bibinfo {author}
  {\bibfnamefont {Q.}~\bibnamefont {Xue}}, \bibinfo {author} {\bibfnamefont
  {J.}~\bibnamefont {Zhang}}, \bibinfo {author} {\bibfnamefont {E.~W.}\
  \bibnamefont {Plummer}}, \bibinfo {author} {\bibfnamefont {X.}~\bibnamefont
  {Zhu}}, \ and\ \bibinfo {author} {\bibfnamefont {J.}~\bibnamefont {Guo}},\
  }\href {\doibase 10.1103/PhysRevB.94.081116} {\bibfield  {journal} {\bibinfo
  {journal} {Phys. Rev. B}\ }\textbf {\bibinfo {volume} {94}},\ \bibinfo
  {pages} {081116} (\bibinfo {year} {2016})}\BibitemShut {NoStop}%
\bibitem [{\citenamefont {Rebec}\ \emph {et~al.}(2017)\citenamefont {Rebec},
  \citenamefont {Jia}, \citenamefont {Zhang}, \citenamefont {Hashimoto},
  \citenamefont {Lu}, \citenamefont {Moore},\ and\ \citenamefont
  {Shen}}]{RebecPRL2017}%
  \BibitemOpen
  \bibfield  {author} {\bibinfo {author} {\bibfnamefont {S.~N.}\ \bibnamefont
  {Rebec}}, \bibinfo {author} {\bibfnamefont {T.}~\bibnamefont {Jia}}, \bibinfo
  {author} {\bibfnamefont {C.}~\bibnamefont {Zhang}}, \bibinfo {author}
  {\bibfnamefont {M.}~\bibnamefont {Hashimoto}}, \bibinfo {author}
  {\bibfnamefont {D.-H.}\ \bibnamefont {Lu}}, \bibinfo {author} {\bibfnamefont
  {R.~G.}\ \bibnamefont {Moore}}, \ and\ \bibinfo {author} {\bibfnamefont
  {Z.-X.}\ \bibnamefont {Shen}},\ }\href {\doibase
  10.1103/PhysRevLett.118.067002} {\bibfield  {journal} {\bibinfo  {journal}
  {Phys. Rev. Lett.}\ }\textbf {\bibinfo {volume} {118}},\ \bibinfo {pages}
  {067002} (\bibinfo {year} {2017})}\BibitemShut {NoStop}%
\bibitem [{\citenamefont {of~cooperative effect on the enhanced superconducting
  transition temperature at~the FeSe/SrTiO3~interface}(2019)}]{SongNC2019}%
  \BibitemOpen
  \bibfield  {author} {\bibinfo {author} {\bibfnamefont {E.}~\bibnamefont
  {of~cooperative effect on the enhanced superconducting transition temperature
  at~the FeSe/SrTiO3~interface}},\ }\href {\doibase 10.1038/s41467-019-08560-z}
  {\bibfield  {journal} {\bibinfo  {journal} {Nature Communications}\ }\textbf
  {\bibinfo {volume} {10}},\ \bibinfo {pages} {758} (\bibinfo {year}
  {2019})}\BibitemShut {NoStop}%
\bibitem [{\citenamefont {Rademaker}\ \emph {et~al.}(2021)\citenamefont
  {Rademaker}, \citenamefont {Alvarez-Suchini}, \citenamefont {Nakatsukasa},
  \citenamefont {Wang},\ and\ \citenamefont {Johnston}}]{RademakerPRB2021}%
  \BibitemOpen
  \bibfield  {author} {\bibinfo {author} {\bibfnamefont {L.}~\bibnamefont
  {Rademaker}}, \bibinfo {author} {\bibfnamefont {G.}~\bibnamefont
  {Alvarez-Suchini}}, \bibinfo {author} {\bibfnamefont {K.}~\bibnamefont
  {Nakatsukasa}}, \bibinfo {author} {\bibfnamefont {Y.}~\bibnamefont {Wang}}, \
  and\ \bibinfo {author} {\bibfnamefont {S.}~\bibnamefont {Johnston}},\ }\href
  {\doibase 10.1103/PhysRevB.103.144504} {\bibfield  {journal} {\bibinfo
  {journal} {Phys. Rev. B}\ }\textbf {\bibinfo {volume} {103}},\ \bibinfo
  {pages} {144504} (\bibinfo {year} {2021})}\BibitemShut {NoStop}%
\bibitem [{\citenamefont {Chandrasekhar}(1962)}]{Chandrasekhar1962}%
  \BibitemOpen
  \bibfield  {author} {\bibinfo {author} {\bibfnamefont {B.~S.}\ \bibnamefont
  {Chandrasekhar}},\ }\href {\doibase 10.1063/1.1777362} {\bibfield  {journal}
  {\bibinfo  {journal} {Applied Physics Letters}\ }\textbf {\bibinfo {volume}
  {1}},\ \bibinfo {pages} {7} (\bibinfo {year} {1962})},\ \Eprint
  {http://arxiv.org/abs/https://doi.org/10.1063/1.1777362}
  {https://doi.org/10.1063/1.1777362} \BibitemShut {NoStop}%
\bibitem [{\citenamefont {Clogston}(1962)}]{ClogstonPRL1962}%
  \BibitemOpen
  \bibfield  {author} {\bibinfo {author} {\bibfnamefont {A.~M.}\ \bibnamefont
  {Clogston}},\ }\href {\doibase 10.1103/PhysRevLett.9.266} {\bibfield
  {journal} {\bibinfo  {journal} {Phys. Rev. Lett.}\ }\textbf {\bibinfo
  {volume} {9}},\ \bibinfo {pages} {266} (\bibinfo {year} {1962})}\BibitemShut
  {NoStop}%
\bibitem [{\citenamefont {Xie}\ \emph {et~al.}(2020)\citenamefont {Xie},
  \citenamefont {Zhou},\ and\ \citenamefont {Law}}]{XiePRL2020}%
  \BibitemOpen
  \bibfield  {author} {\bibinfo {author} {\bibfnamefont {Y.-M.}\ \bibnamefont
  {Xie}}, \bibinfo {author} {\bibfnamefont {B.~T.}\ \bibnamefont {Zhou}}, \
  and\ \bibinfo {author} {\bibfnamefont {K.~T.}\ \bibnamefont {Law}},\ }\href
  {\doibase 10.1103/PhysRevLett.125.107001} {\bibfield  {journal} {\bibinfo
  {journal} {Phys. Rev. Lett.}\ }\textbf {\bibinfo {volume} {125}},\ \bibinfo
  {pages} {107001} (\bibinfo {year} {2020})}\BibitemShut {NoStop}%
\bibitem [{\citenamefont {Sau}\ \emph {et~al.}(2010)\citenamefont {Sau},
  \citenamefont {Lutchyn}, \citenamefont {Tewari},\ and\ \citenamefont
  {Das~Sarma}}]{sauPRL2010}%
  \BibitemOpen
  \bibfield  {author} {\bibinfo {author} {\bibfnamefont {J.~D.}\ \bibnamefont
  {Sau}}, \bibinfo {author} {\bibfnamefont {R.~M.}\ \bibnamefont {Lutchyn}},
  \bibinfo {author} {\bibfnamefont {S.}~\bibnamefont {Tewari}}, \ and\ \bibinfo
  {author} {\bibfnamefont {S.}~\bibnamefont {Das~Sarma}},\ }\href {\doibase
  10.1103/PhysRevLett.104.040502} {\bibfield  {journal} {\bibinfo  {journal}
  {Phys. Rev. Lett.}\ }\textbf {\bibinfo {volume} {104}},\ \bibinfo {pages}
  {040502} (\bibinfo {year} {2010})}\BibitemShut {NoStop}%
\bibitem [{\citenamefont {Marsiglio}(2020)}]{MARSIGLIO2020168102}%
  \BibitemOpen
  \bibfield  {author} {\bibinfo {author} {\bibfnamefont {F.}~\bibnamefont
  {Marsiglio}},\ }\href {\doibase https://doi.org/10.1016/j.aop.2020.168102}
  {\bibfield  {journal} {\bibinfo  {journal} {Annals of Physics}\ }\textbf
  {\bibinfo {volume} {417}},\ \bibinfo {pages} {168102} (\bibinfo {year}
  {2020})},\ \bibinfo {note} {eliashberg theory at 60: Strong-coupling
  superconductivity and beyond}\BibitemShut {NoStop}%
\bibitem [{sup()}]{supple}%
  \BibitemOpen
  \href@noop {} {\bibinfo  {journal} {See Supplemental Material at
  http://****** for the detailed methodology and more results with other
  parameters.}\ }\BibitemShut {NoStop}%
\bibitem [{\citenamefont {Fukui}\ \emph {et~al.}(2005)\citenamefont {Fukui},
  \citenamefont {Hatsugai},\ and\ \citenamefont {Suzuki}}]{FukuiJPSJ2005}%
  \BibitemOpen
\bibfield  {journal} {  }\bibfield  {author} {\bibinfo {author} {\bibfnamefont
  {T.}~\bibnamefont {Fukui}}, \bibinfo {author} {\bibfnamefont
  {Y.}~\bibnamefont {Hatsugai}}, \ and\ \bibinfo {author} {\bibfnamefont
  {H.}~\bibnamefont {Suzuki}},\ }\href {\doibase 10.1143/JPSJ.74.1674}
  {\bibfield  {journal} {\bibinfo  {journal} {Journal of the Physical Society
  of Japan}\ }\textbf {\bibinfo {volume} {74}},\ \bibinfo {pages} {1674}
  (\bibinfo {year} {2005})}\BibitemShut {NoStop}%
\bibitem [{\citenamefont {Bardeen}\ \emph {et~al.}(1957)\citenamefont
  {Bardeen}, \citenamefont {Cooper},\ and\ \citenamefont
  {Schrieffer}}]{BCS1957}%
  \BibitemOpen
  \bibfield  {author} {\bibinfo {author} {\bibfnamefont {J.}~\bibnamefont
  {Bardeen}}, \bibinfo {author} {\bibfnamefont {L.~N.}\ \bibnamefont {Cooper}},
  \ and\ \bibinfo {author} {\bibfnamefont {J.~R.}\ \bibnamefont {Schrieffer}},\
  }\href {\doibase 10.1103/PhysRev.108.1175} {\bibfield  {journal} {\bibinfo
  {journal} {Phys. Rev.}\ }\textbf {\bibinfo {volume} {108}},\ \bibinfo {pages}
  {1175} (\bibinfo {year} {1957})}\BibitemShut {NoStop}%
\bibitem [{\citenamefont {Li}\ and\ \citenamefont
  {Johnston}(2020)}]{Shaozhinpj}%
  \BibitemOpen
  \bibfield  {author} {\bibinfo {author} {\bibfnamefont {S.}~\bibnamefont
  {Li}}\ and\ \bibinfo {author} {\bibfnamefont {S.}~\bibnamefont {Johnston}},\
  }\href {\doibase 10.1038/s41535-020-0243-3} {\bibfield  {journal} {\bibinfo
  {journal} {npj Quantum Materials}\ }\textbf {\bibinfo {volume} {5}},\
  \bibinfo {pages} {40} (\bibinfo {year} {2020})}\BibitemShut {NoStop}%
\bibitem [{\citenamefont {Kitaev}(2003)}]{KITAEV20032}%
  \BibitemOpen
  \bibfield  {author} {\bibinfo {author} {\bibfnamefont {A.}~\bibnamefont
  {Kitaev}},\ }\href {\doibase https://doi.org/10.1016/S0003-4916(02)00018-0}
  {\bibfield  {journal} {\bibinfo  {journal} {Annals of Physics}\ }\textbf
  {\bibinfo {volume} {303}},\ \bibinfo {pages} {2} (\bibinfo {year}
  {2003})}\BibitemShut {NoStop}%
\bibitem [{\citenamefont {Nayak}\ \emph
  {et~al.}(2008{\natexlab{b}})\citenamefont {Nayak}, \citenamefont {Simon},
  \citenamefont {Stern}, \citenamefont {Freedman},\ and\ \citenamefont
  {Das~Sarma}}]{NayakRMP2008}%
  \BibitemOpen
  \bibfield  {author} {\bibinfo {author} {\bibfnamefont {C.}~\bibnamefont
  {Nayak}}, \bibinfo {author} {\bibfnamefont {S.~H.}\ \bibnamefont {Simon}},
  \bibinfo {author} {\bibfnamefont {A.}~\bibnamefont {Stern}}, \bibinfo
  {author} {\bibfnamefont {M.}~\bibnamefont {Freedman}}, \ and\ \bibinfo
  {author} {\bibfnamefont {S.}~\bibnamefont {Das~Sarma}},\ }\href {\doibase
  10.1103/RevModPhys.80.1083} {\bibfield  {journal} {\bibinfo  {journal} {Rev.
  Mod. Phys.}\ }\textbf {\bibinfo {volume} {80}},\ \bibinfo {pages} {1083}
  (\bibinfo {year} {2008}{\natexlab{b}})}\BibitemShut {NoStop}%
\bibitem [{\citenamefont {Sarma}\ \emph {et~al.}(2015)\citenamefont {Sarma},
  \citenamefont {Freedman},\ and\ \citenamefont {Nayak}}]{sarma2015}%
  \BibitemOpen
  \bibfield  {author} {\bibinfo {author} {\bibfnamefont {S.~D.}\ \bibnamefont
  {Sarma}}, \bibinfo {author} {\bibfnamefont {M.}~\bibnamefont {Freedman}}, \
  and\ \bibinfo {author} {\bibfnamefont {C.}~\bibnamefont {Nayak}},\ }\href
  {https://www.nature.com/articles/npjqi20151} {\bibfield  {journal} {\bibinfo
  {journal} {Nature News}\ } (\bibinfo {year} {2015})}\BibitemShut {NoStop}%
\bibitem [{\citenamefont {Klitzing}\ \emph {et~al.}(1980)\citenamefont
  {Klitzing}, \citenamefont {Dorda},\ and\ \citenamefont
  {Pepper}}]{KlitzingPRL1980}%
  \BibitemOpen
  \bibfield  {author} {\bibinfo {author} {\bibfnamefont {K.~v.}\ \bibnamefont
  {Klitzing}}, \bibinfo {author} {\bibfnamefont {G.}~\bibnamefont {Dorda}}, \
  and\ \bibinfo {author} {\bibfnamefont {M.}~\bibnamefont {Pepper}},\ }\href
  {\doibase 10.1103/PhysRevLett.45.494} {\bibfield  {journal} {\bibinfo
  {journal} {Phys. Rev. Lett.}\ }\textbf {\bibinfo {volume} {45}},\ \bibinfo
  {pages} {494} (\bibinfo {year} {1980})}\BibitemShut {NoStop}%
\end{thebibliography}%
\end{document}